\documentclass[12pt]{iopart}
\usepackage{iopams}

\usepackage[pdftex]{graphicx}
\usepackage{bm}% bold math
\usepackage{overpic}
\usepackage{xcolor}
\usepackage[numbers,sort&compress]{natbib}
\newcommand{\eqref}[1]{(\ref{#1})}
\newcommand{\dd}{\mathrm{d}}
\begin{document}

\title{Lagrangian velocity statistics of homogeneous isotropic turbulence in dilute polymer solutions}

\author{Yusuke Koide \& Susumu Goto}

\address{Graduate School of Engineering Science, The University of Osaka, 1-3 Machikaneyama, Toyonaka, Osaka 560-8531, Japan}
\ead{y.koide.es@osaka-u.ac.jp}
\vspace{10pt}
% \begin{indented}
% \item[]August 2017
% \end{indented}

\begin{abstract}
We conduct direct numerical simulations of homogeneous isotropic turbulence in dilute polymer solutions to investigate the Lagrangian velocity statistics.
We show how polymers modulate the power spectral density of the Lagrangian velocity and the Lagrangian integral timescale by varying the Reynolds number, forcing method, and polymer relaxation time.
As the polymer relaxation time increases, the attenuation of the power spectral density extends successively from high to low frequencies, and the Lagrangian integral timescale increases.
To clarify the mechanism underlying the modulation of the Lagrangian velocity statistics, we decompose the Lagrangian velocity into the contributions from vortices at different length scales.
Using this scale-decomposition analysis, we demonstrate that the observed modulation of the Lagrangian velocity statistics results from polymer-induced suppression of vortices that proceeds from smaller to larger scales.
\end{abstract}

%\maketitle

\section{Introduction}

Polymers have received considerable attention as drag-reducing agents for many decades.
High-molecular-weight polymers suppress small-scale vortices in turbulence, thereby reducing friction drag by up to $80\%$~\cite{Virk1975-qv,White2008-gb,Xi2019-nc}.
Owing to its remarkable effectiveness, drag reduction using polymer additives has been applied to oil transport~\cite{Burger1982-dy}, firefighting~\cite{Fabula,Figueredo2003-vt}, and irrigation~\cite{Bouchenafa2021-hx}.
To support and expand industrial applications, it is essential to understand how polymers modulate turbulent flows.

Accordingly, numerous studies have investigated the Eulerian statistics of polymer turbulence, including the mean velocity, the Reynolds stress, and the energy spectrum.
For instance, the mean velocity profile in channel flow exhibits a complex variation depending on the drag reduction rate, while it approaches a universal function irrespective of the polymer type and concentration at the maximum drag reduction limit~\cite{Virk1967-rv,Warholic1999-mz,Ptasinski2001-ux,Min2003-qn}.
Polymers also qualitatively modulate the spatial structures of turbulence.
Several experiments and numerical simulations have reported that the slope of the energy spectrum $E(k)$ becomes steeper upon the addition of polymers~\cite{Perlekar2010-rn,Vonlanthen2013-rb,Valente2016-xz,Yamani2021-eu,Rosti2023-fk,Singh2025-ur}.
Rosti \textit{et al.}~\cite{Rosti2023-fk} conducted large-scale direct numerical simulations (DNS) of dilute polymer solutions at high Reynolds and Deborah numbers, where the Deborah number $\mathrm{De}$ is defined as the ratio of the polymer relaxation time to the eddy-turnover time of the largest-scale eddies.
They showed that, with increasing $\mathrm{De}$, the elastic scaling regime characterized by $E(k)\propto k^{-2.3}$ extends from high wave numbers and spans the entire range at $\mathrm{De}\simeq 1$.
A further increase in $\mathrm{De}$ caused $E(k)$ to revert to the Kolmogorov scaling.
Zhang \textit{et al.}~\cite{Zhang2021-zh} experimentally observed that the elastic scaling regime emerged in the second-order longitudinal velocity structure function.
Thus, rather than merely suppressing turbulence, polymers induce non-trivial chaotic flows coupled with elasticity, known as elasto-inertial turbulence~\cite{Samanta2013-xu,Dubief2023-ye}.

In addition to the Eulerian statistics, the Lagrangian statistics play an important role in the fundamental physics of turbulence modulation by polymers.
Polymer dynamics, which are essential to turbulence modulation, are strongly governed by the Lagrangian flow history~\cite{Vincenzi2007-vs,Benzi2018-jy,Kumar2023-mm}.
Polymer stretching and alignment in turbulence have been widely investigated using Brownian dynamics simulations of bead-spring chains and dumbbell models~\cite{Massah1993-lw,Stone2003-sa,Zhou2003-gk,Terrapon2004-xb,Watanabe2010-oi,Picardo2023-ok,Koide2024-lc}.
Terrapon \textit{et al.}~\cite{Terrapon2004-xb} added the dumbbell model into a Newtonian turbulent channel flow.
Using conditional statistics, they identified the flow history relevant to the dumbbell stretching.
Watanabe and Gotoh~\cite{Watanabe2010-oi} conducted Brownian dynamics simulations of the finitely extensible nonlinear elastic~(FENE) dumbbells in Newtonian isotropic turbulence and demonstrated the dynamical relationship between the dumbbell stretching and the velocity gradient tensor along the Lagrangian trajectory.
Koide and Goto~\cite{Koide2024-lc} revealed the stretching mechanism of dumbbells by different-scale vortices in turbulence by analyzing the dynamical properties of the scale-decomposed velocity gradient from the Lagrangian viewpoint.

Beyond their influence on polymer dynamics, the Lagrangian properties of turbulence are indispensable because they control the transport of passive scalars, such as temperature and mass~\cite{Yeung2002-xy}.
Several studies have investigated the polymer effects on heat and mass transfer~\cite{Matthys1991-gu,Gupta2005-ax,Gasljevic2007-io,Vaithianathan2007-gi,Benzi2010-xm,Ahlers2010-cv,Rahul2026-wj}.
Gupta \textit{et al.}~\cite{Gupta2005-ax} examined the passive scalar transport in viscoelastic turbulent channel flows with DNS.
They demonstrated that when a significant drag reduction occurred, the stabilization of the low-speed streaks caused an increase in the streamwise heat flux, whereas the wall-normal and spanwise heat fluxes decreased due to the suppression of turbulent fluctuations in these directions.
In addition, it has been reported that polymers can influence the clustering of inertial particles by modulating turbulent flows~\cite{De-Lillo2012-py,Sinhuber2018-cl}.

Despite these nontrivial transport processes induced by polymers, the Lagrangian statistics in turbulent flows with polymer additives remain only partially understood.
Crawford \textit{et al.}~\cite{Crawford2008-dn} conducted experimental measurements of the Lagrangian acceleration in turbulent flows of polymer solutions and observed an increase in the Lagrangian acceleration autocorrelation time.
de Chaumont Quitry and Ouellette~\cite{de-Chaumont-Quitry2016-xf} experimentally measured the Lagrangian velocity structure function in dilute polymer solutions and observed two distinct master curves depending on the polymer concentration.
However, systematic investigations of the Lagrangian statistics for different parameters, such as the polymer relaxation time and the Reynolds number, are still lacking.
Consequently, while polymer effects on the Eulerian statistics, including the energy spectrum and the second-order structure function, are well established~\cite{Valente2016-xz,Yamani2021-eu,Zhang2021-zh,Rosti2023-fk,Singh2025-ur}, their Lagrangian counterparts~(i.e., the power spectral density of the Lagrangian velocity $E_L(\omega)$ and the second-order Lagrangian structure function $S_2(\tau)$) remain unclear.

In the present study, we aim to reveal how polymers modulate the Lagrangian properties of turbulence by conducting DNS of dilute polymer solutions.
For this purpose, we employ the finitely extensible nonlinear elastic dumbbell model with the Peterlin's approximation~(FENE-P)~\cite{bird1987dynamics}, which captures the elasticity and shear-thinning behavior of dilute polymer solutions.
This study focuses on statistically steady turbulence in a periodic cube sustained by the forcing to elucidate the fundamental aspects of the Lagrangian statistics in dilute polymer solutions.
By varying the polymer relaxation time, we demonstrate how polymer elasticity alters the power spectral density of the Lagrangian velocity and the Lagrangian integral timescale.
Furthermore, we show the relation between the modulations of the Lagrangian statistics and the spatial structures of turbulence by using a scale-decomposition analysis of the Lagrangian velocity.

\section{Methods}
\label{sec:method}

\subsection{Direct numerical simulations}
We conduct DNS of turbulence in dilute polymer solutions described by the FENE-P model.
The simulations employ periodic boundary conditions in three orthogonal directions with period $2\pi$.
The velocity field $\bm{u}(\bm{x},t)$ and the pressure field $p(\bm{x},t)$ follow 
\begin{equation}
  \nabla\cdot \bm{u} = 0\label{eq:continuity}
\end{equation}
and
\begin{equation}
  \rho\left\{\frac{\partial \bm{u}}{\partial t} +(\bm{u}\cdot \nabla)\bm{u}\right\}= - \nabla p +\mu_{\mathrm{s}}\nabla^2\bm{u} + \nabla\cdot \bm{\sigma}_{\mathrm{p}}+\bm{f},\label{eq:momenta}
\end{equation}
where $\rho$ is the fluid density, $\mu_{\mathrm{s}}$ is the solvent viscosity, $\bm{\sigma}_{\mathrm{p}}(\bm{x},t)$ is the polymer contribution to the stress tensor, and $\bm{f}(\bm{x},t)$ is the external force. 
In the FENE-P model, $\bm{\sigma}_{\mathrm{p}}$ is given by
\begin{equation}
  \bm{\sigma}_{\mathrm{p}} = \frac{\mu_{\mathrm{p}}}{\tau}\left\{f(\bm{C})\bm{C}-\bm{I}\right\},~f(\bm{C})=\frac{L_\mathrm{p}^2-3}{L_\mathrm{p}^2-\mathrm{tr}\bm{C}},
\end{equation}
where $\bm{C}(\bm{x},t)=\langle \bm{R}\bm{R}\rangle/(R_0^2/3) $ is the conformation tensor with $\bm{R}$ being the end-to-end vector of polymers and $R_0^2$ being the mean square end-to-end distance of polymers in equilibrium, $\mu_{\mathrm{p}}$ is the polymer contribution to the zero-shear viscosity, $\tau$ is the polymer relaxation time, $L_\mathrm{p}$ is the maximum extensibility of polymers, and $\bm{I}$ is the identity tensor.
Here, $\bm{C}(\bm{x},t)$ obeys
\begin{equation}
  \frac{\partial \bm{C}}{\partial t} +(\bm{u}\cdot \nabla )\bm{C}-\bm{C}\cdot (\nabla \bm{u}) - (\nabla\bm{u})^\mathsf{T}\cdot \bm{C} = -\frac{1}{\tau} \left\{f(\bm{C})\bm{C}-\bm{I}\right\}.\label{eq:conformation}
\end{equation}

The statistically steady turbulence is sustained by two types of external forces.
The first is the deterministic force $\bm{f}^\mathrm{(D)}(\bm{x},t)$~\cite{Lamorgese2005-mf} whose Fourier coefficient $\widehat{\bm{f}}^{(\mathrm{D})}(\bm{k},t)$ is expressed as 
\begin{equation}
\label{eq:cases_f}
\widehat{\bm{f}}^{(\mathrm{D})}(\bm{k},t)=
\left\{
\begin{array}{ll}
\displaystyle
\frac{P}{2E_{k_f}(t)}\widehat{\bm{u}}(\bm{k},t)
& 0<|\bm{k}|\leq k_f, \\[6pt]
\bm{0}
& \mathrm{otherwise},
\end{array}
\right.
\end{equation}
where $P$ is the energy input rate, $k_f(=2.5)$ is the maximum forcing wavenumber, $\widehat{\bm{u}}(\bm{k},t)$ is the Fourier coefficient of $\bm{u}(\bm{x},t)$, and $E_{k_f}(t)$ is defined as 
\begin{equation}
    E_{k_f}(t) = \sum_{0<|\bm{k}|\leq k_f}\frac{1}{2}|\widehat{\bm{u}}(\bm{k},t)|^2.
\end{equation}
The second is the stochastic force $\bm{f}^\mathrm{(S)}(\bm{x},t)$~\cite{Alvelius1999-cr}.
We impose $\bm{f}^\mathrm{(S)}(\bm{x},t)$ on the velocity field in the low-wavenumber range centered at $k_f=2$.
The forcings by both $\bm{f}^\mathrm{(D)}$ and $\bm{f}^\mathrm{(S)}$ lead to a constant energy input rate $P$, which is set to unity in this study.

We employ the Simplified Marker and Cell~(SMAC) method to numerically integrate Eqs.~\eqref{eq:continuity} and \eqref{eq:momenta}; we use the first-order Euler method for the pressure term, the second-order Adams--Bashforth method for the convection and viscous terms, and the trapezoidal rule for the polymer stress term with the spatial derivatives evaluated by the second-order central finite difference.
We apply the algorithm proposed by Vaithianathan \textit{et al.}~\cite{Vaithianathan2006-bf} to numerically integrate Eq.~\eqref{eq:conformation}.

\subsection{Tracking of fluid particles}
To obtain the Lagrangian velocity, we track $64^3$ fluid particles whose initial positions are uniformly distributed.
The position $\bm{x}_L(t|\bm{x}_0,t_0)$ of each fluid particle obeys 
\begin{equation}
  \frac{\dd}{\dd t}\bm{x}_L(t|\bm{x}_0,t_0) = \bm{v}_L(t|\bm{x}_0,t_0),\label{eq:fluid_particle}
\end{equation}
where $\bm{v}_L(t|\bm{x}_0,t_0)$ is the Lagrangian velocity, $t_0$ is the labeling time, and $\bm{x}_0$ is the position at $t=t_0$.
The Lagrangian velocity $\bm{v}_L(t|\bm{x}_0,t_0)$ is related to the Eulerian velocity $\bm{u}(\bm{x},t)$ by 
\begin{equation}
  \bm{v}_L(t|\bm{x}_0,t_0) = \bm{u}(\bm{x}_L(t|\bm{x}_0,t_0),t).
\end{equation}
We use trilinear interpolation for the fluid velocity $\bm{u}(\bm{x}_L(t|\bm{x}_0,t_0),t)$ and integrate Eq.~\eqref{eq:fluid_particle} using the second-order Adams--Bashforth method. 

\subsection{Parameters}
To systematically evaluate the effect of polymers on the Lagrangian velocity, we conduct DNS for various values of the Reynolds number and the polymer relaxation time using two types of forcing methods.
Table~\ref{table:Parameter_newtonian} shows the DNS parameters and statistics of Newtonian turbulence considered:
$N^3$ is the number of grid points; $\mathrm{Re}_\lambda$ is the Reynolds number based on the Taylor microscale $\lambda$; $\Delta x$ is the grid width; $\eta=(\nu^{3}/\overline{\epsilon})^{1/4}$ is the Kolmogorov length, where $\nu$ is the kinematic viscosity and $\overline{\epsilon}=2\nu \overline{S_{ij}S_{ij}}$ is the mean dissipation rate of kinetic energy per unit mass, with $\bm{S}=[\nabla\bm{u}+\left(\nabla\bm{u}\right)^\mathsf{T}]/2$.
Here, $\overline{(\cdot)}$ denotes the spatio-temporal average.
We obtain $\mathrm{Re}_\lambda$ as
\begin{equation}
    \mathrm{Re}_\lambda=\sqrt{\frac{20}{3\nu\overline{\epsilon}}}K,
\end{equation}
where $K$ is the kinetic energy per unit mass.
The Courant--Friedrichs--Lewy~(CFL) number is defined as $\sqrt{2K/3}\Delta t/\Delta x$, where $\Delta t$ is the time step.

For each Newtonian case listed in Table~\ref{table:Parameter_newtonian}, we perform DNS of viscoelastic turbulence described by the FENE-P model, using the same $\bm{f}$, $N^3$, and $\nu$.
Note that for the FENE-P model, $\nu=\nu_{\mathrm{s}}+\nu_{\mathrm{p}}$ is defined as the sum of the solvent contribution $\nu_{\mathrm{s}}$ and the polymer contribution $\nu_{\mathrm{p}}$ in the zero-shear limit.
In presenting the results, we use the subscript $0$ to denote reference quantities of the corresponding Newtonian case. 
Thus, $\mathrm{Re}_{\lambda,0}$ denotes the Reynolds number based on the Taylor microscale of the corresponding Newtonian case.
In the present study, we change the degree of viscoelasticity through $\tau$ while fixing the values of $\beta=\mu_{\mathrm{s}}/(\mu_{\mathrm{s}}+\mu_{\mathrm{p}})$ and $L_\mathrm{p}$ at $0.9$ and $1000$, respectively.
Table~\ref{table:Parameter_fenep} lists the DNS parameters and statistics for the FENE-P cases considered.
The Weissenberg number is defined as $\mathrm{Wi}_\eta=\tau/\tau_{\eta,0}$, where $\tau_{\eta,0}$ is the Kolmogorov time of the corresponding Newtonian case.
We define the Eulerian large-eddy turnover time $T_{E}$ as $L/u^\prime$, where $u^\prime$ denotes the root-mean-square of the velocity fluctuation and $L$ is the integral length given by 
\begin{equation}
  L = \frac{3\pi}{4K}\int_0^\infty k^{-1}E(k)\dd k,
\end{equation}
where $E(k)$ denotes the energy spectrum.
The Deborah number is then defined as $\mathrm{De}=\tau/T_{E,0}$, where $T_{E,0}$ is the Eulerian large-eddy turnover time of the corresponding Newtonian case.
We have verified that the spatial resolution is sufficient to evaluate the statistics of the Lagrangian velocity~(see Appendix~A).

% --------------------
\begin{table}
  \caption{Parameters and statistics of Newtonian turbulence: $\bm{f}$ is the external force, $N^3$ is the number of grid points, $\mathrm{Re}_\lambda$ is the Reynolds number based on the Taylor microscale $\lambda$, $\eta$ is the Kolmogorov length, and $\Delta x$ is the grid width.
  The Courant--Friedrichs--Lewy~(CFL) number is defined as $\sqrt{2K/3}\Delta t/\Delta x$, where $K$ is the kinetic energy per unit mass and $\Delta t$ is the time step. }
  \label{table:Parameter_newtonian}  
  \begin{center}
  \begin{tabular}{cccccc}
  \hline
  &$\bm{f}$ & $N^3$&$\mathrm{Re}_\lambda$&$\eta/\Delta x$&CFL number\\
  \hline
  Run384D&$\bm{f}^\mathrm{(D)} $&$384^3$ &$135$ & $0.68$&$4.0\times 10^{-2}$ \\
  Run512D&$\bm{f}^\mathrm{(D)} $&$512^3$ &$218$  & $0.46$&$5.4\times 10^{-2}$ \\
  Run512S&$\bm{f}^\mathrm{(S)} $&$512^3$ &$225$  & $0.46$&$5.5\times 10^{-2}$ \\
  \hline
  \end{tabular}
  \end{center}
  \end{table}
% --------------------
% --------------------
\begin{table}
  \centering
  \caption{Parameters and statistics of viscoelastic turbulence described by the FENE-P model: $\mathrm{Wi}_\eta=\tau/\tau_{\eta,0}$ is the Weissenberg number based on the Kolmogorov time $\tau_{\eta,0}$ of the corresponding Newtonian case; $\mathrm{De}=\tau/T_{E,0}$ is the Deborah number with $T_{E,0}$ being the Eulerian large-eddy turnover time of the corresponding Newtonian case; $\beta=\mu_{\mathrm{s}}/(\mu_{\mathrm{s}}+\mu_{\mathrm{p}})$ is the ratio of the solvent viscosity to the zero-shear viscosity of the solution; $L_\mathrm{p}$ is the maximum extensibility of polymers.
}
  \label{table:Parameter_fenep}  
  % \begin{center}
  (a) Run384D\\
  \begin{tabular}{cccc}
  \hline
  $\mathrm{Wi}_\eta$&$\mathrm{De}$&$\beta$&$L_p$\\
  \hline
  % $0.5$&$0.028$ &$0.9$ & $1000$\\
  $1$&$0.056$ &$0.9$ & $1000$\\
  $3$&$0.17$ &$0.9$ & $1000$\\
  $6$&$0.34$ &$0.9$ & $1000$\\
  $15$&$0.84$ &$0.9$ & $1000$\\
  \hline
  \end{tabular}
  % \end{center}
    % \begin{center}

  (b) Run512D\\
  \begin{tabular}{cccc}
  \hline
  $\mathrm{Wi}_\eta$&$\mathrm{De}$&$\beta$&$L_p$\\
  \hline
  $1$&$0.036$ &$0.9$ & $1000$\\
  $3$&$0.11$ &$0.9$ & $1000$\\
  $6$&$0.22$ &$0.9$ & $1000$\\
  $15$&$0.54$ &$0.9$ & $1000$\\
  \hline

  \end{tabular}
  % \end{center}
    % \begin{center}
  
    (c) Run512S\\
  \begin{tabular}{cccc}
  \hline
  $\mathrm{Wi}_\eta$&$\mathrm{De}$&$\beta$&$L_p$\\
  \hline
  $1$&$0.043$ &$0.9$ & $1000$\\
  $3$&$0.13$ &$0.9$ & $1000$\\
  $6$&$0.26$ &$0.9$ & $1000$\\
  $15$&$0.65$ &$0.9$ & $1000$\\
  \hline

  \end{tabular}
  % \end{center}
  \end{table}
% --------------------
\section{Results and Discussion}
\label{sec:results}

\subsection{Turbulence modulation by polymers from the Eulerian viewpoint}

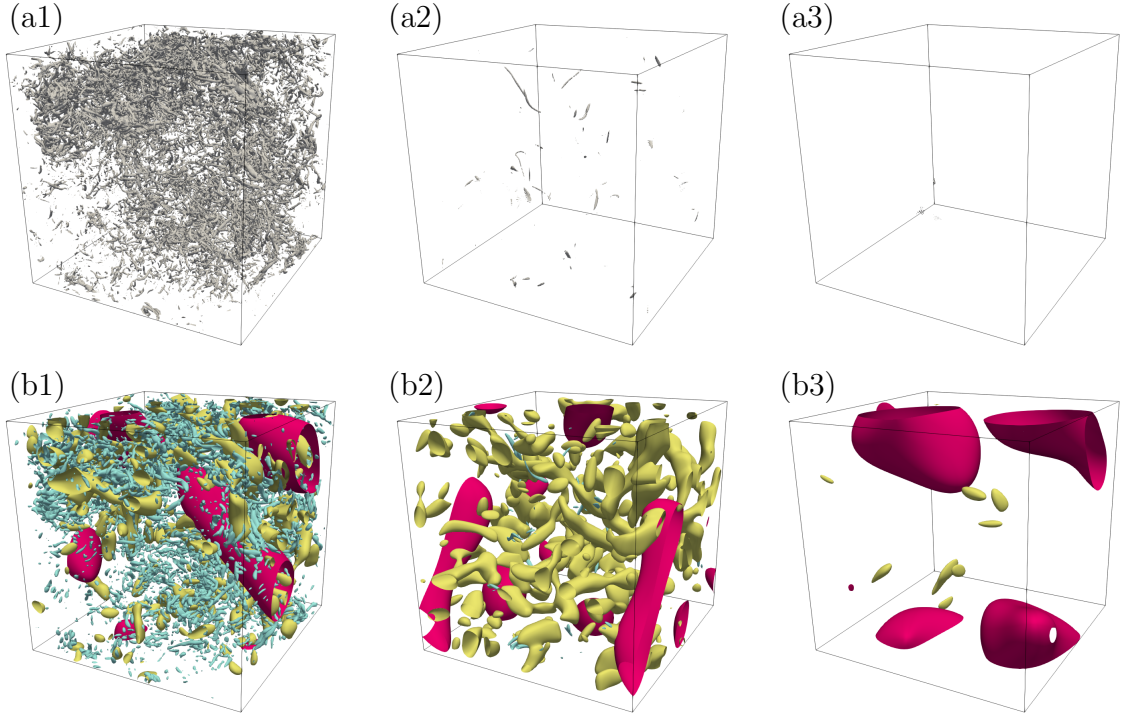
\begin{figure}
  \centering
  \begin{overpic}[width=1\linewidth]{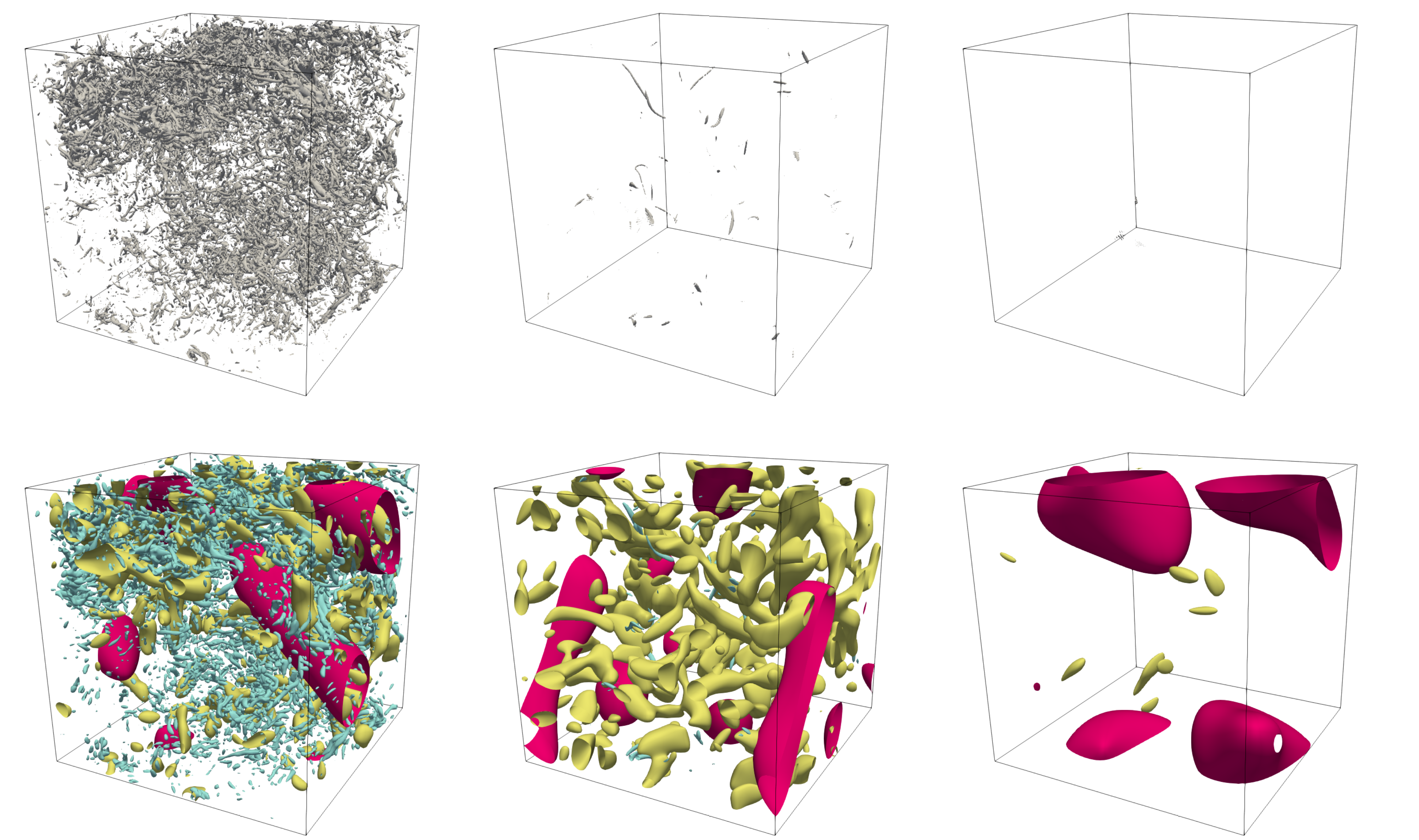} 
        \put(2,58){(a1)}
        \put(2,27){(b1)}
        \put(34,58){(a2)}
        \put(34,27){(b2)}
        \put(67,58){(a3)}
        \put(67,27){(b3)}
  \end{overpic}
    \caption{Isosurfaces of (a) raw $|\bm{\omega}|^2$ and (b) bandpass-filtered enstrophy $|\bm{\omega}^{(k_c)}|^2$ for $\mathrm{Re}_{\lambda,0}=218$ with the forcing $\bm{f}^{(\mathrm{D})}$ (Run512D) at (1) $\mathrm{Wi}_\eta=0$~(i.e., Newtonian turbulence), (2) $6$, and (3) $15$.
    In (a), the threshold for the isosurface is set to $\mu +3\sigma$, where $\mu$ and $\sigma$ denote the spatio-temporal average and standard deviation of $|\bm{\omega}|^2$ in the Newtonian case. 
    In (b), different colors indicate different values of $k_c$: red, $k_c=\sqrt{2}$; yellow, $4\sqrt{2}$; cyan, $16\sqrt{2}$.
    The threshold for the isosurface is set to $\mu^{(k_c)} +2\sigma^{(k_c)}$ for $k_c=\sqrt{2}$ and $4\sqrt{2}$, and to $\mu^{(k_c)} +3\sigma^{(k_c)}$ for $k_c=16\sqrt{2}$, where $\mu^{(k_c)}$ and $\sigma^{(k_c)}$ denote the spatio-temporal average and standard deviation of $|\bm{\omega}^{(k_c)}|^2$ in the Newtonian case.}
  \label{fig:snapshot}
\end{figure}

This subsection focuses on the modulation of turbulence structures and Eulerian statistics induced by polymers before presenting the Lagrangian velocity statistics.
Figure~\ref{fig:snapshot}(a) shows isosurfaces of the enstrophy $|\bm{\omega}|^2$ for $\mathrm{Re}_{\lambda,0}=218$ with the forcing $\bm{f}^{(\mathrm{D})}$ (Run512D) at different values of the Weissenberg number $\mathrm{Wi}_\eta$.
For all cases, the threshold for the isosurfaces is set to $\mu+3\sigma$, where $\mu$ and $\sigma$ denote the spatio-temporal average and standard deviation of $|\bm{\omega}|^2$ in the Newtonian case.
A comparison of Figs.~\ref{fig:snapshot}(a1--a3) demonstrates that vortical structures are suppressed by polymers~\cite{Perlekar2010-rn,Ur_Rehman2022-ra,Rosti2023-fk}.
Note that $|\bm{\omega}|^2$, a quantity based on the velocity gradient, is dominated by the smallest-scale vortices.
To extract vortices at different length scales in viscoelastic fluids, we apply scale decomposition to the velocity field $\bm{u}(\bm{x},t)$.
Specifically, we define the velocity $\bm{u}^{(k_c)}(\bm{x},t)$ associated with the wavenumber $k_c$ as the bandpass-filtered velocity with the passband $k_c/\sqrt{2}\leq k<\sqrt{2}k_c$~\cite{Goto2017-im,Hirota2020-jm}.
We then focus on the bandpass-filtered enstrophy $|\bm{\omega}^{(k_c)}|^2=|\nabla\times \bm{u}^{(k_c)}|^2$.
Figure~\ref{fig:snapshot}(b) presents the isosurfaces of $|\bm{\omega}^{(k_c)}|^2$ for $k_c=\sqrt{2}$, $4\sqrt{2}$, and $16\sqrt{2}$~($k_c\eta_0=0.0080$, $0.032$, and $0.13$, respectively, where $\eta_0$ is the Kolmogorov length of the Newtonian case) at the same instant as the corresponding panels in Fig.~\ref{fig:snapshot}(a).
Here, we set the threshold for the isosurfaces using the spatio-temporal average $\mu^{(k_c)}$ and standard deviation $\sigma^{{(k_c)}}$ of $|\bm{\omega}^{(k_c)}|^2$ in the Newtonian case.
Figure~\ref{fig:snapshot}(b1) demonstrates that vortices at different length scales coexist hierarchically in the Newtonian turbulence.
We observe in Fig.~\ref{fig:snapshot}(b2) that, for $\mathrm{Wi}_\eta=6$, vortices at $k_c=16\sqrt{2}$ almost disappear, while vortices at $k_c=\sqrt{2}$ and $4\sqrt{2}$ are similar to those in the Newtonian case.
For $\mathrm{Wi}_\eta=15$, vortices at $k_c=4\sqrt{2}$ also become weak as well as those at $k_c=16\sqrt{2}$, whereas the largest-scale vortices at $k_c=\sqrt{2}$ still exist.
Thus, visualization using the bandpass filter qualitatively demonstrates that as the polymer relaxation time increases, vortex attenuation extends from smaller to larger scales~\cite{Masuda2026submitted}.
%   --------------------
%   --------------------

\begin{figure}
  \centering
  \begin{overpic}[width=0.5\linewidth]{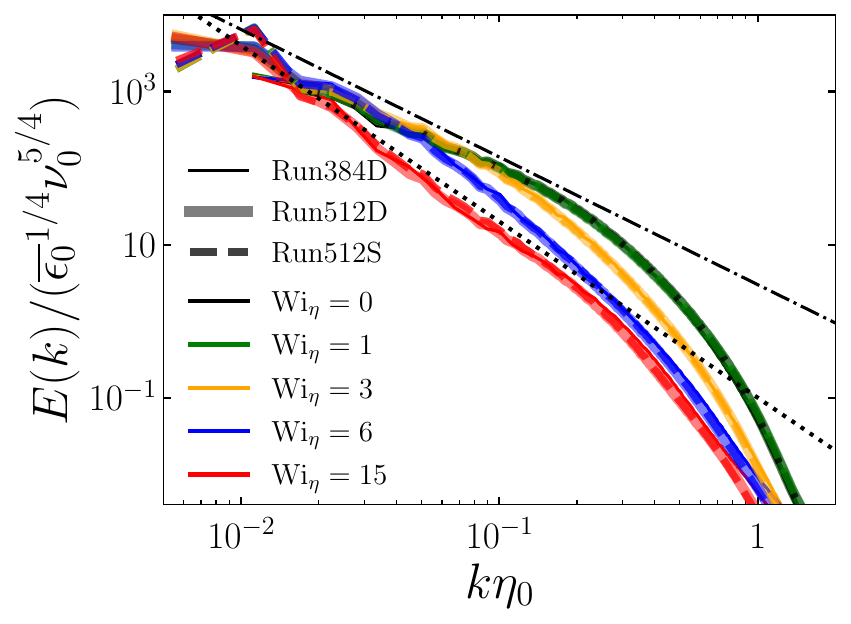} 
  \end{overpic}
  \caption{Energy spectrum $E(k)$ normalized by $\overline{\epsilon_0}^{1/4}\nu_0^{5/4}$ as a function of $k\eta_0$ for Run384D, Run512D, and Run512S at different values of $\mathrm{Wi}_\eta$. The black dash-dotted and dotted lines indicate $E(k)\propto k^{-5/3}$ and $E(k)\propto k^{-2.3}$, respectively.}
  \label{fig:ene_spe}
\end{figure}

We then investigate the energy spectrum $E(k)$ to quantitatively evaluate the effect of polymers on the spatial structure of turbulence.
Figure~\ref{fig:ene_spe} shows $E(k)$ for the various parameters listed in Table~\ref{table:Parameter_fenep}.
We normalize $k$ and $E(k)$ using the kinematic viscosity $\nu_0$ and the mean energy dissipation rate $\overline{\epsilon_0}$ of the corresponding Newtonian case to compare data for different $\mathrm{Re}_{\lambda,0}$, $\bm{f}$, and $\mathrm{Wi}_\eta$.
We also note that, in the present simulations, the zero-shear kinematic viscosity of each FENE-P case is set equal to that of the corresponding Newtonian case, i.e., $\nu=\nu_{\mathrm{s}}+\nu_{\mathrm{p}}=\nu_0$.
For the FENE-P model, the mean dissipation rate of kinetic energy per unit mass is written as $\overline{\epsilon}=\overline{\epsilon_{\mathrm{s}}}+\overline{\epsilon_{\mathrm{p}}}$, where $\overline{\epsilon_{\mathrm{s}}}=2\nu_{\mathrm{s}}\overline{S_{ij}S_{ij}}$ denotes the solvent viscous contribution and $\overline{\epsilon_{\mathrm{p}}}=\overline{\sigma_{\mathrm{p},ij}S_{ij}}/\rho$ denotes the polymer contribution, which represents the mean rate of work done by the flow against the polymer stress~\cite{de-Angelis2005-fn,Dallas2010-za,Valente2014-aq}.
Because the turbulence is statistically steady and the forcing terms $\bm{f}^{(\mathrm{D})}$ and $\bm{f}^{(\mathrm{S})}$ are designed so that the energy input rate $P$ is constant in time and equal to unity for all cases, $\overline{\epsilon}=\overline{\epsilon_0}=P$.
The present normalization is not intended to match the smallest scales based on the solvent viscosity $\nu_{\mathrm{s}}$ and the solvent viscous dissipation rate $\overline{\epsilon_{\mathrm{s}}}$.
Rather, it is chosen to collapse $E(k)$ in the large-scale part of the inertial range, where the flows are only weakly modified by polymers.
As shown below, this normalization clarifies both the wavenumber range where the polymer-induced suppression occurs and the extent to which $E(k)$ is reduced relative to the Newtonian case.

To examine the $\mathrm{Wi}_\eta$ dependence, we focus on $E(k)$ for Run512D~(thick solid lines), which corresponds to $\mathrm{Re}_{\lambda,0}=218$ with the forcing $\bm{f}^{(\mathrm{D})}$.
As $\mathrm{Wi}_\eta$ increases, $E(k)$ decreases from the high-wavenumber range relative to that for Newtonian fluids.
This observation is consistent with the visualization shown in Fig.~\ref{fig:snapshot} and previous studies~\cite{Perlekar2010-rn,Rosti2023-fk,Masuda2026submitted}. 
The preferential suppression of smaller-scale fluctuations can be attributed to their shorter characteristic timescales, or equivalently higher strain rates, which lead to stronger interactions with polymers~\cite{lumley1969drag,Tabor1986-yi}.
For $\mathrm{Wi}_\eta=15$, $E(k)$ is suppressed throughout the inertial range and exhibits a power law $E(k)\propto k^{-2.3}$, as reported in previous studies~\cite{Rosti2023-fk,Singh2025-ur}.
Although $E(k)$ deviates from the Newtonian spectrum with increasing $\mathrm{Wi}_\eta$, its low-wavenumber component remains almost unchanged, which is also consistent with the visualization of $|\bm{\omega}^{(k_c)}|^2$ shown in Fig.~\ref{fig:snapshot}.
For a more comprehensive description of the Eulerian statistics, we refer to the time series of the Eulerian velocity $\bm{u}_E(t|\bm{x})$ at a fixed position $\bm{x}$.
We confirm that the power spectral density of $\bm{u}_E(t|\bm{x})$ almost coincides with the one-dimensional energy spectrum for both Newtonian and viscoelastic fluids, indicating that $\bm{u}_E(t|\bm{x})$ reflects spatial information about turbulence due to the sweeping effect by large-scale flows~(see Appendix~B).

Let us now consider the effects of $\mathrm{Re}_{\lambda,0}$ and $\bm{f}$.
Figure~\ref{fig:ene_spe} demonstrates that for fixed $\mathrm{Wi}_\eta$, $E(k)$ almost collapses onto a single function in the high-wavenumber range regardless of $\mathrm{Re}_{\lambda,0}$ and $\bm{f}$.
This collapse indicates that large-scale flows have little influence on the polymer-induced suppression of small-scale vortices.
In the next subsection, we demonstrate how the Lagrangian statistics are modulated in these turbulent flows of dilute polymer solutions.

%   --------------------
%   --------------------

\subsection{Statistics of the Lagrangian velocity}

\begin{figure}
  \centering
  \begin{overpic}[width=1\linewidth]{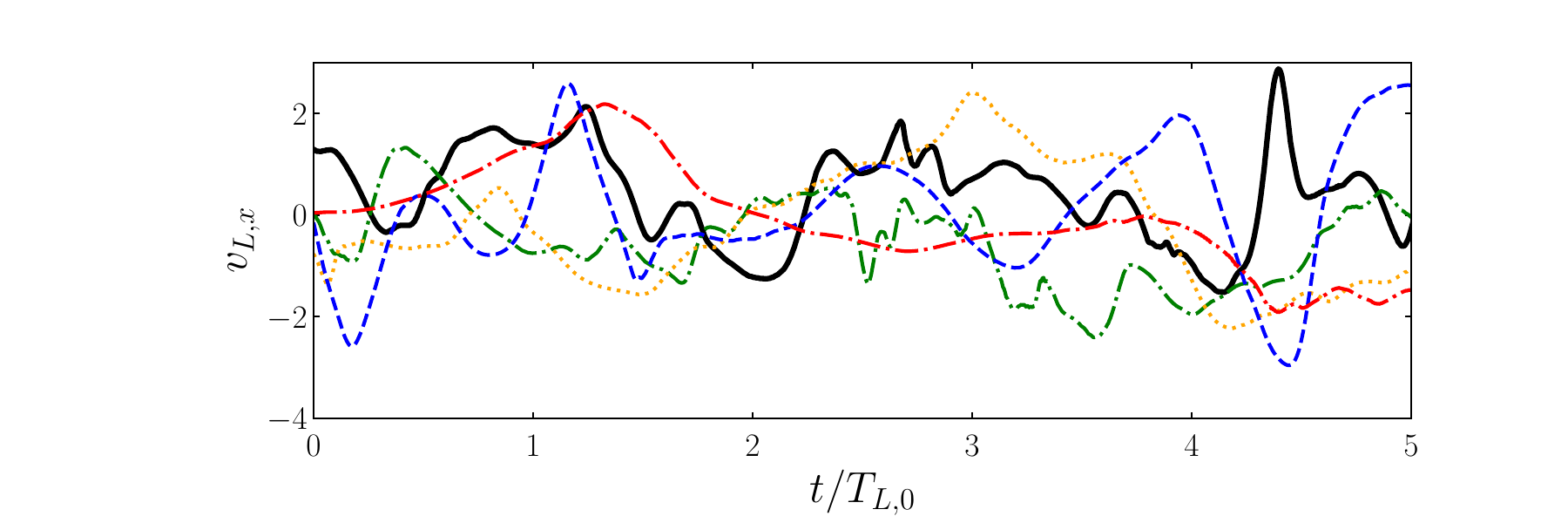} 
  \end{overpic}
  \caption{Time series of the $x$-component of the Lagrangian velocity $v_{L,x}(t|\bm{x}_0,t_0)$ as a function of the normalized time $t/T_{L,0}$ for $\mathrm{Re}_{\lambda,0}=218$ with the forcing $\bm{f}^{(\mathrm{D})}$ (Run512D) at $\mathrm{Wi}_\eta=0$~(black solid line), $1$~(green dash-dotted line), $3$~(orange dotted line), $6$~(blue dashed line), and $15$~(red dash-dotted line).}
  \label{fig:lag_vel}
\end{figure}

This subsection focuses on the effect of polymers on the Lagrangian velocity.
Figure~\ref{fig:lag_vel} shows the time series of the Lagrangian velocity $v_{L,x}(t|\bm{x}_0,t_0)$ in the $x$ direction for $\mathrm{Re}_{\lambda,0}=218$ with the forcing $\bm{f}^{(\mathrm{D})}$ (Run512D) at various $\mathrm{Wi}_\eta$, where $t$ is normalized by the Lagrangian integral timescale $T_{L,0}$ for the Newtonian case.
Here, we define the Lagrangian integral timescale $T_{L}$ of $\bm{v}_{L}(t|\bm{x}_0,t_0)$ as 
\begin{equation}
  T_{L}=\int_0^\infty C_{L}(t)\dd t,\label{eq:Lag_integ}
\end{equation} 
where the autocorrelation function $C_{L}(t)$ is given by
\begin{equation}
    C_{L}(t) = \frac{\langle \bm{v}_{L}(s+t|\bm{x}_0,t_0)\cdot \bm{v}_{L}(s|\bm{x}_0,t_0)\rangle}{\langle |\bm{v}_{L}|^2\rangle }
\end{equation}
with $\langle \cdot \rangle$ denoting the ensemble average over $64^3$ trajectories.
Figure~\ref{fig:lag_vel} shows that the temporal fluctuations of $v_{L,x}(t|\bm{x}_0,t_0)$ become slower as $\mathrm{Wi}_\eta$ increases.
%   --------------------
%   --------------------

\begin{figure*}
  \centering
      \begin{tabular}{c}
      \begin{minipage}{0.33\hsize}
          \begin{overpic}[width=1\linewidth]{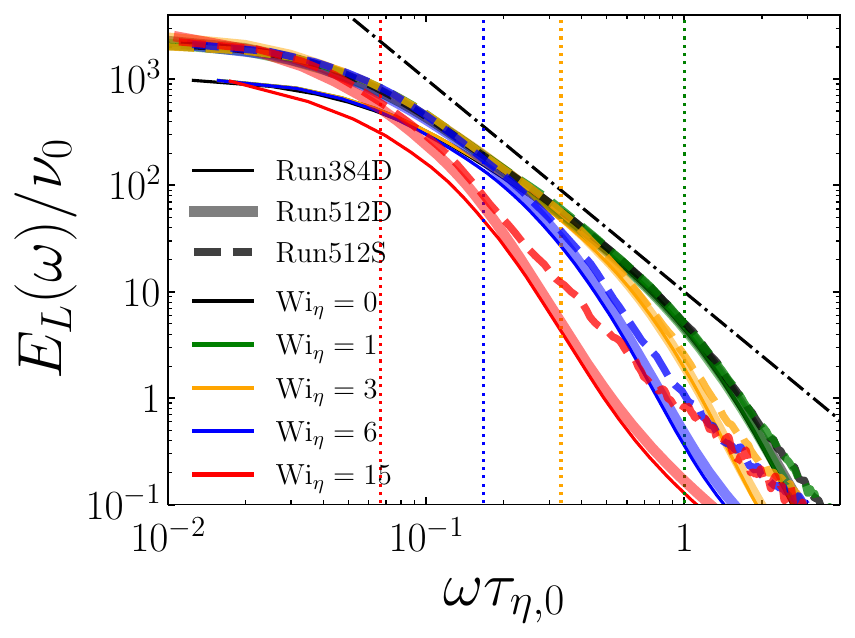}
              \linethickness{3pt}
        \put(5,70){(a)}

          \end{overpic}
      \end{minipage}
      \begin{minipage}{0.33\hsize}
          \begin{overpic}[width=1\linewidth]{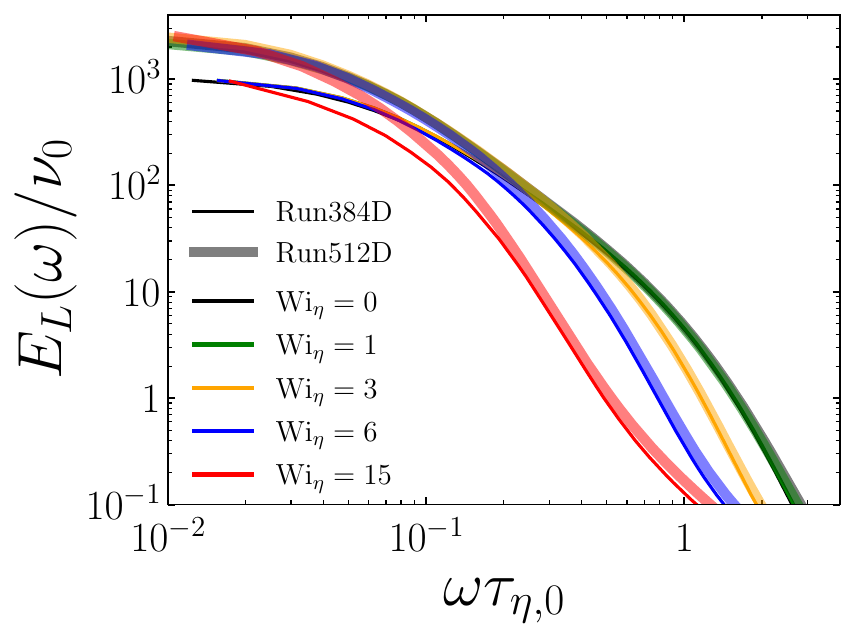}
              \linethickness{3pt}
        \put(5,73){(b)}

          \end{overpic}
      \end{minipage}
          \begin{minipage}{0.33\hsize}
          \begin{overpic}[width=1\linewidth]{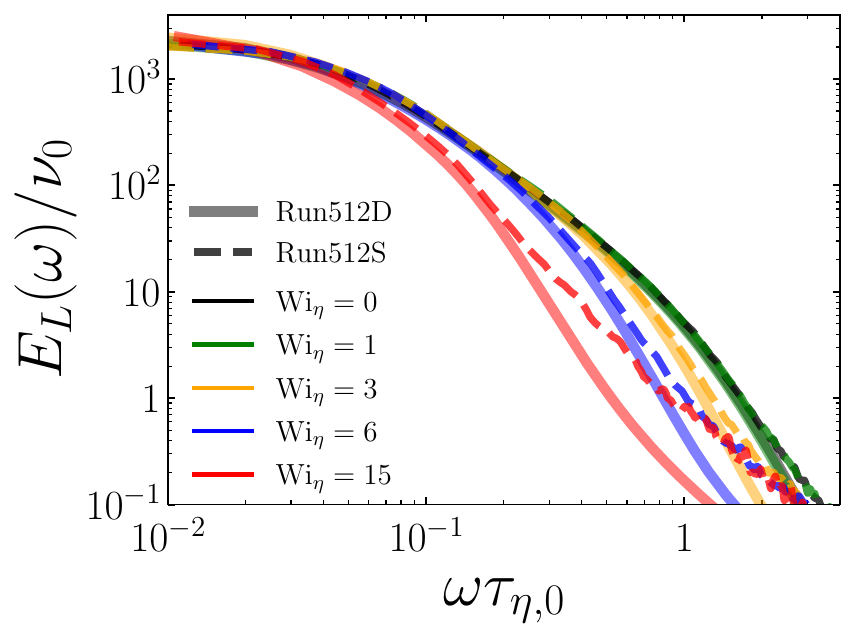}
              \linethickness{3pt}
        \put(5,73){(c)}

          \end{overpic}
      \end{minipage}
      \end{tabular}
      \caption{Power spectral density $E_{L}(\omega)$ of the Lagrangian velocity $\bm{v}_{L}(t|\bm{x}_0,t_0)$ normalized by $\nu_0$ as a function of $\omega\tau_{\eta,0}$ at different values of $\mathrm{Wi}_\eta$: (a) Run384D, Run512D, and Run512S; (b) Run384D and Run512D; (c) Run512D and Run512S. The black dash-dotted line in (a) indicates $E_{L}(\omega)\propto \omega^{-2}$. The dotted vertical lines in (a) indicate $\omega=1/\tau$~(i.e., $\omega\tau_{\eta,0}=1/\mathrm{Wi}_\eta$), with the colors corresponding to the values of $\mathrm{Wi}_\eta$.}
      \label{fig:lag_psd}
\end{figure*}

To quantify this polymer effect in the frequency domain, we then investigate the power spectral density $E_L(\omega)$ of $\bm{v}_{L}(t|\bm{x}_0,t_0)$.
In this study, we define $E_{L}(\omega)$ as
\begin{equation}
    E_{L}(\omega) = \frac{2\langle|\hat{\bm{v}}_{L}(\omega)|^2\rangle}{T}, \label{eq:Lagrangian_power_spectra}
\end{equation}
where $\hat{\bm{v}}_{L}(\omega)$ is the Fourier component of $\bm{v}_{L}(t|\bm{x}_0,t_0)$ and $T$ is the length of the time series of $\bm{v}_{L}(t|\bm{x}_0,t_0)$.
Note that $E_{L}(\omega)$ defined in Eq.~\eqref{eq:Lagrangian_power_spectra} satisfies 
\begin{equation}
    \int_0^\infty E_{L}(\omega) \dd\omega = \langle \bm{v}_{L}^2\rangle.%\lim_{T\to \infty}\frac{1}{T}\int_0^T v_{L,\alpha}^2dt.
\end{equation}
Figure~\ref{fig:lag_psd}(a) presents $E_{L}(\omega)$ for the same parameters as in Fig.~\ref{fig:ene_spe}.
As in Fig.~\ref{fig:ene_spe} for $E(k)$, $\omega$ and $E_{L}(\omega)$ are normalized using $\nu_0$ and $\overline{\epsilon_0}$.
For Newtonian turbulence, the scaling law $E_{L}(\omega)\propto  \omega^{-2}$ is known to hold in the Lagrangian inertial range~\cite{Yeung2001-rq,Mordant2001-td}, as expected by the Kolmogorov similarity hypothesis.
Although $\mathrm{Re}_{\lambda,0}$ considered in this study is not high enough for a clear Lagrangian inertial range to emerge~\cite{Yeung2006-nl}, Fig.~\ref{fig:lag_psd}(a) shows a weak indication of the scaling $E_L(\omega)\propto\omega^{-2}$ for $\mathrm{Wi}_\eta=0$~(i.e., Newtonian turbulence).
To investigate the $\mathrm{Wi}_\eta$ dependence, let us focus on the results for Run512D~(thick solid lines), which corresponds to $\mathrm{Re}_{\lambda,0}=218$ with the forcing $\bm{f}^{(\mathrm{D})}$.
As $\mathrm{Wi}_\eta$ increases, $E_{L}(\omega)$ in the high-frequency range decreases, consistent with Fig.~\ref{fig:lag_vel}, leading to the disappearance of the Lagrangian inertial range. 
In contrast, $E_{L}(\omega)$ in the low-frequency range almost coincides with that for the Newtonian case.
Thus, we demonstrate that the attenuation of $E_{L}(\omega)$ extends successively from high to low frequencies with increasing $\mathrm{Wi}_\eta$.
While this study focuses solely on the relaxation time among polymer parameters, we have confirmed that moderate changes in $\beta$ and $L_\mathrm{p}$ have negligible effects on $E(k)$ and $E_L(\omega)$~(see Appendix C).
To provide a reference for interpreting the modulation behavior of $E_L(\omega)$, we indicate $\omega=1/\tau$ (i.e., $\omega\tau_{\eta,0}=1/\mathrm{Wi}_\eta$) by vertical dotted lines in Fig.~\ref{fig:lag_psd}(a).
For $\mathrm{Wi}_\eta=3$ and $6$, the modulation of $E_L(\omega)$ appears mainly in the range $\omega\tau_{\eta,0}\gtrsim 1/\mathrm{Wi}_\eta$, whereas this correspondence is less clear for $\mathrm{Wi}_\eta=1$ and $15$.
Thus, the polymer relaxation time provides only a rough guide to the frequency range affected by polymers.
In the next subsection, we use a scale-decomposition analysis to examine how the modulation of $E_L(\omega)$ is related to the changes in the hierarchy of coherent vortices shown in Fig.~\ref{fig:snapshot}.

Next, to examine the Reynolds-number dependence, we compare $E_{L}(\omega)$ for $\mathrm{Re}_{\lambda,0}=135$~(Run384D; thin solid lines) and $\mathrm{Re}_{\lambda,0}=218$~(Run512D; thick solid lines) at fixed $\mathrm{Wi}_\eta$~[see Fig.~\ref{fig:lag_psd}(b)].
For fixed $\mathrm{Wi}_\eta$ and $\bm{f}$, $E_{L}(\omega)$ for different $\mathrm{Re}_{\lambda,0}$ collapses onto a single curve in the high-frequency range.
This collapse indicates that the modulation of the Lagrangian velocity is governed primarily by $\mathrm{Wi}_\eta$, i.e., the ratio of the polymer relaxation time to the timescale of the smallest-scale vortices.
It is worth noting that although we focus on $\mathrm{Wi}_\eta$ to characterize polymer relaxation time, the Deborah number $\mathrm{De}$~(i.e., the ratio of the polymer relaxation time to the timescale of the largest-scale flow) is also an important dimensionless number.
Rosti \textit{et al.}~\cite{Rosti2023-fk} demonstrated that once $\mathrm{De}$ exceeds unity, $E(k)$ again exhibits the Kolmogorov scaling.
Thus, for $\mathrm{De}\gtrsim 1$, both $\mathrm{De}$ and $\mathrm{Wi}_\eta$ might collectively influence $E_{L}(\omega)$, an issue that needs to be addressed in future work.

Finally, to investigate the effect of the forcing method, we compare $E_{L}(\omega)$ for $\bm{f}^{(\mathrm{D})}$~(Run512D; thick solid lines) and $\bm{f}^{(\mathrm{S})}$~(Run512S; dashed lines), which have similar values of $\mathrm{Re}_{\lambda,0}$~[see Fig.~\ref{fig:lag_psd}(c)].
For fixed $\mathrm{Wi}_\eta$, $E_{L}(\omega)$ is almost independent of $\bm{f}$, except in the high-frequency range, where the deviation becomes more pronounced with increasing $\mathrm{Wi}_\eta$.
As demonstrated in our previous study~\cite{Koide2025-lz}, $E_{L}(\omega)$ of Newtonian turbulence is more sensitive to the forcing method than $E(k)$, leading to non-universal behavior of $E_{L}(\omega)$.
As shown in Sec.~\ref{subsec:decomp}, even in the FENE-P cases, the larger values of $E_{L}(\omega)$ in the high-frequency range for $\bm{f}^\mathrm{(S)}$ compared with $\bm{f}^\mathrm{(D)}$ are attributed to the effect of the largest-scale flows, whose relative influence increases due to the suppression of small-scale vortices at large $\mathrm{Wi}_\eta$.
It should be noted that our results do not include the effect of mean flows because $\bm{f}^\mathrm{(D)}$ and $\bm{f}^\mathrm{(S)}$ generate turbulent flows without mean flows.
Given that mean flows, or the largest-scale flows driven by external forcing, were found to affect $E_L(\omega)$ over a broad frequency range in Newtonian turbulence~\cite{Koide2025-lz}, clarifying the role of mean flows in the Lagrangian properties of turbulence in dilute polymer solutions remains an important problem for future study.

% %   --------------------
%   --------------------

\begin{figure*}
  \centering
      \begin{tabular}{c}
      \begin{minipage}{0.5\hsize}
          \begin{overpic}[width=1\linewidth]{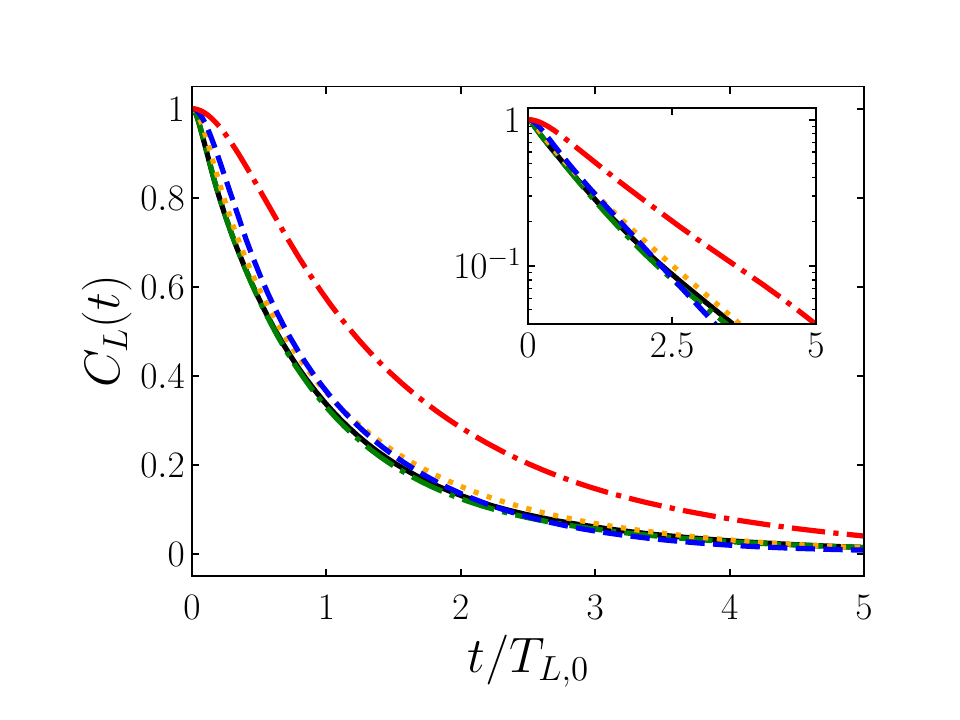}
              \linethickness{3pt}
        \put(5,63){(a)}

          \end{overpic}
      \end{minipage}
      \begin{minipage}{0.5\hsize}
          \begin{overpic}[width=1\linewidth]{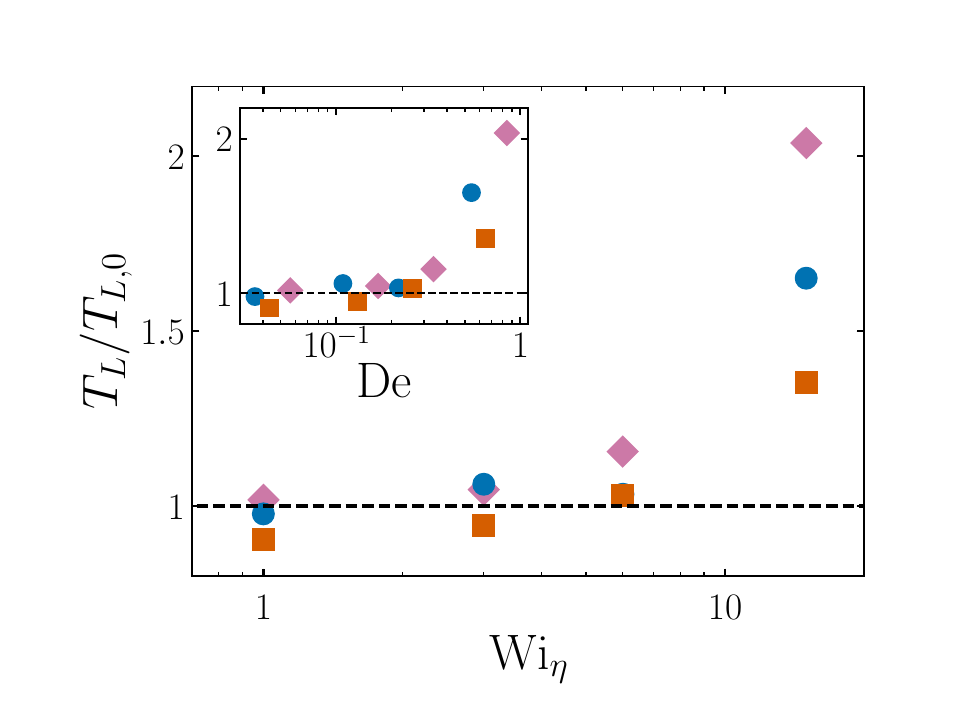}
              \linethickness{3pt}
        \put(5,63){(b)}

          \end{overpic}
      \end{minipage}
      \end{tabular}
      \caption{(a) Autocorrelation function $C_L(t)$ of the Lagrangian velocity $\bm{v}_{L}(t|\bm{x}_0,t_0)$ as a function of $t/T_{L,0}$ for $\mathrm{Re}_{\lambda,0}=218$ with the forcing $\bm{f}^{(\mathrm{D})}$ (Run512D) at $\mathrm{Wi}_\eta=0$~(black solid), $1$~(green dash-dotted), $3$~(orange dotted), $6$~(blue dashed), and $15$~(red dash-dotted). The inset shows a semi-logarithmic plot of $C_L(t)$. (b) Lagrangian integral timescale $T_{L}$ normalized by $T_{L,0}$ as a function of the Weissenberg number $\mathrm{Wi}_\eta$ for Run384D~(diamond), Run512D~(circle), and Run512S~(square). The dashed line indicates $T_{L}/T_{L,0}=1$. The inset shows $T_{L}/T_{L,0}$ as a function of the Deborah number $\mathrm{De}$.}
      \label{fig:lag_acf}
\end{figure*}

Before closing this subsection, we also refer to the autocorrelation function $C_L(t)$ of the Lagrangian velocity.
Although $C_L(t)$ corresponds to the Fourier transform of $E_{L}(\omega)$, $C_L(t)$ provides direct insight into the temporal history of $\bm{v}_{L}(t|\bm{x}_0,t_0)$ in the time domain.
Figure~\ref{fig:lag_acf}(a) shows $C_L(t)$ for $\mathrm{Re}_{\lambda,0}=218$ with the forcing $\bm{f}^{(\mathrm{D})}$ (Run512D) at different values of $\mathrm{Wi}_\eta$.
Here, $t$ is normalized by the Lagrangian integral timescale $T_{L,0}$ for Newtonian turbulence~[Eq.~\eqref{eq:Lag_integ}].
We observe that regardless of $\mathrm{Wi}_\eta$, $C_L(t)$ exhibits an exponential decay, except at short times~[see the inset of Fig.~\ref{fig:lag_acf}(a)].
For $\mathrm{Wi}_\eta\leq 6$, the decay of $C_L(t)$ for $t/T_{L,0}\lesssim 1$ becomes slightly slower with increasing $\mathrm{Wi}_\eta$, while $C_L(t)$ for $t/T_{L,0}\gtrsim 1$ remains almost unaffected.
For $\mathrm{Wi}_\eta=15$, $C_L(t)$ decays more slowly than in the other cases over the entire range.
To systematically demonstrate the $\mathrm{Wi}_\eta$ dependence of the correlation in the Lagrangian velocity, Fig.~\ref{fig:lag_acf}(b) shows $T_L$ normalized by $T_{L,0}$ as a function of $\mathrm{Wi}_\eta$ for all cases.
We observe that $T_L$ increases with $\mathrm{Wi}_\eta$ regardless of $\mathrm{Re}_{\lambda,0}$ and $\bm{f}$, indicating that polymers with longer relaxation times prolong the memory of the Lagrangian velocity.
This tendency is consistent with a previous experiment by Crawford \textit{et al.}~\cite{Crawford2008-dn}, which reported an increase in the autocorrelation time of the Lagrangian acceleration in dilute polymer solutions.
Although $T_L/T_{L,0}$ scatters for fixed $\mathrm{Wi}_\eta$, the inset of Fig.~\ref{fig:lag_acf}(b) demonstrates that the Deborah number $\mathrm{De}$ is more suitable for describing the increasing tendency of $T_L/T_{L,0}$.
It is reasonable that $\mathrm{De}$ is relevant to $T_L$ because $T_L$ is mainly governed by the largest-scale flows rather than by the smallest-scale vortices, as discussed in the next subsection.

%   --------------------
%   --------------------
\subsection{Scale-decomposition analysis of the Lagrangian velocity}
\label{subsec:decomp}

\begin{figure}
  \centering
  \begin{overpic}[width=1\linewidth]{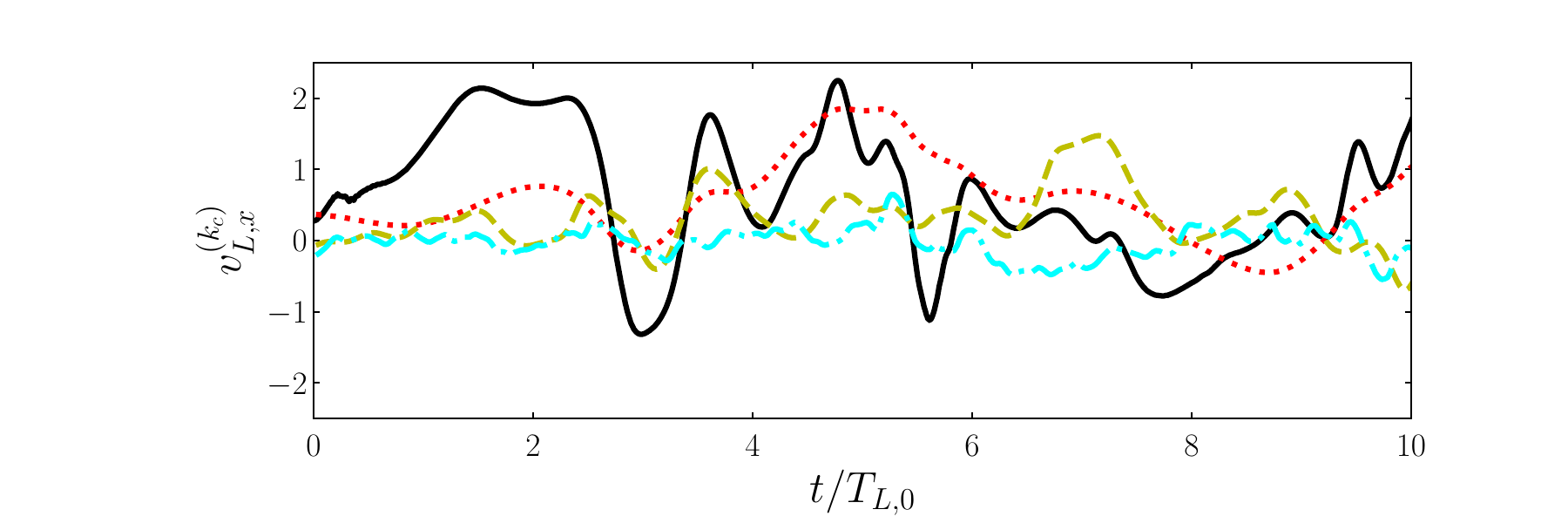} 
  \end{overpic}
  \caption{Time series of the $x$-component of the scale-decomposed Lagrangian velocity $v_{L,x}^{(k_c)}(t|\bm{x}_0,t_0)$ as a function of the normalized time $t/T_{L,0}$ for $\mathrm{Re}_{\lambda,0}=218$ with the forcing $\bm{f}^{(\mathrm{D})}$ (Run512D) at $\mathrm{Wi}_\eta=6$. The results are shown for $k_c=\sqrt{2}(\simeq k_f)$~(red dotted line), $4\sqrt{2}$~(yellow dashed line), and $16\sqrt{2}$~(cyan dash-dotted line), corresponding to $k_c\eta_0=0.0080,\,0.032$, and $0.13$, respectively. The black solid line shows the $x$-component of the raw Lagrangian velocity $v_{L,x}(t|\bm{x}_0,t_0)$.}

  \label{fig:lag_bp_vel}
\end{figure}

The observed modulation of the Lagrangian velocity statistics, including $E_{L}(\omega)$ and $T_L$, reflects changes in the temporal motions experienced by fluid particles and is expected to originate from polymer-induced changes in vortical motions at different scales.
To clarify this relation, we adopt a scale-decomposition analysis of the Lagrangian velocity, which allows us to identify the vortex scales responsible for the polymer-induced changes in Lagrangian velocity statistics.
Specifically, we employ the scale-decomposition method for the Lagrangian velocity based on the bandpass-filtered velocity $\bm{u}^{(k_c)}(\bm{x},t)$~\cite{Koide2025-lz}.
Using $\bm{u}^{(k_c)}(\bm{x},t)$, we define the Lagrangian velocity $\bm{v}_L^{(k_c)}(t|\bm{x}_0,t_0)$ associated with the wavenumber $k_c$ as
\begin{equation}
    \bm{v}_L^{(k_c)}(t|\bm{x}_0,t_0) = \bm{u}^{(k_c)}(\bm{x}_L(t|\bm{x}_0,t_0),t).\label{eq:decomposed_Lagrangian}
\end{equation}
In Eq.~\eqref{eq:decomposed_Lagrangian}, we compute $\bm{u}^{(k_c)}(\bm{x}_L(t|\bm{x}_0,t_0),t)$ by applying trilinear interpolation to $\bm{u}^{(k_c)}(\bm{x},t)$.
Since $\bm{u}(\bm{x}_L(t|\bm{x}_0,t_0),t)=\sum_{k_c}\bm{u}^{(k_c)}(\bm{x}_L(t|\bm{x}_0,t_0),t)$,
Eq.~\eqref{eq:decomposed_Lagrangian} allows us to decompose $\bm{v}_L(t|\bm{x}_0,t_0)(=\bm{u}(\bm{x}_L(t|\bm{x}_0,t_0),t))$ into the contribution from each scale:
\begin{equation}
    \bm{v}_L(t|\bm{x}_0,t_0) = \sum_{k_c} \bm{v}_L^{(k_c)}(t|\bm{x}_0,t_0). \label{eq:Lagrangian_velocity_decomposition}
\end{equation}
Note that the scale decomposition has no effect on the trajectory $\bm{x}_L(t|\bm{x}_0,t_0)$ of fluid particles; the time evolution of $\bm{x}_L(t|\bm{x}_0,t_0)$ is governed by the raw velocity field $\bm{u}(\bm{x},t)$.
Figure~\ref{fig:lag_bp_vel} shows the time series of $v_{L,x}^{(k_c)}(t|\bm{x}_0,t_0)$ for various $k_c$ in Run512D at $\mathrm{Wi}_\eta=6$, where $\mathrm{Re}_{\lambda,0}=218$ and the forcing is $\bm{f}^{(\mathrm{D})}$.
We observe that $v_{L,x}^{(k_c)}(t|\bm{x}_0,t_0)$ at lower $k_c$ tends to exhibit larger-amplitude and slower fluctuations.
%   --------------------
%   --------------------

\begin{figure*}
  \centering
      \begin{tabular}{c}
      \begin{minipage}{0.5\hsize}
          \begin{overpic}[width=1\linewidth]{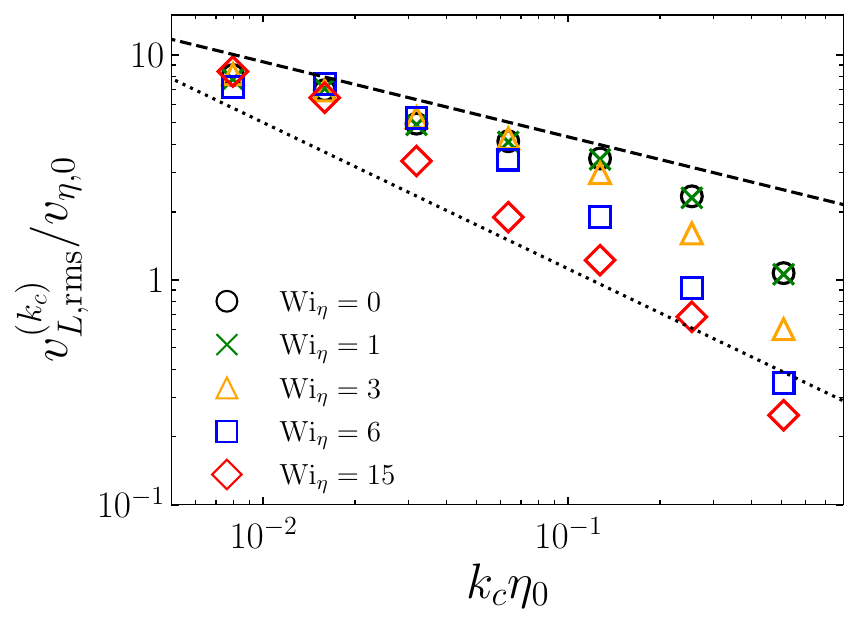}
              \linethickness{3pt}
        \put(5,70){(a)}

          \end{overpic}
      \end{minipage}
      \begin{minipage}{0.5\hsize}
          \begin{overpic}[width=0.98\linewidth]{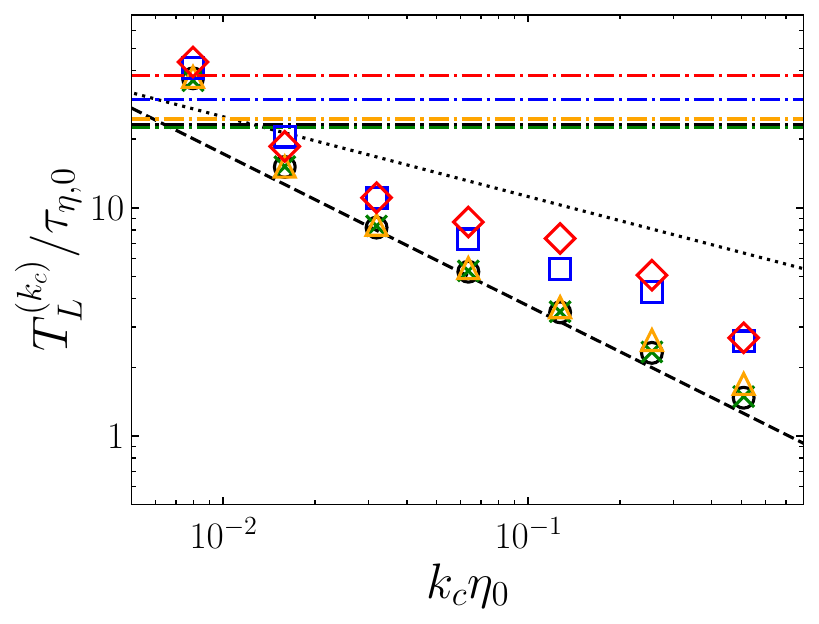}
              \linethickness{3pt}
        \put(5,73){(b)}

          \end{overpic}
      \end{minipage}
      \end{tabular}
      \caption{(a) Root-mean-square amplitude $v_{L,\mathrm{rms}}^{(k_c)}$ and (b) Lagrangian integral timescale $T_L^{(k_c)}$ of the scale-decomposed Lagrangian velocity $\bm{v}_{L}^{(k_c)}(t|\bm{x}_0,t_0)$ as functions of $k_c\eta_0$ for $\mathrm{Re}_{\lambda,0}=218$ with the forcing $\bm{f}^{(\mathrm{D})}$ (Run512D) at various values of $\mathrm{Wi}_\eta$. Here, $v_{L,\mathrm{rms}}^{(k_c)}$ and $T_L^{(k_c)}$ are normalized by $v_{\eta,0}$ and $\tau_{\eta,0}$, respectively. In (a), the dashed and dotted lines indicate $v_{L,\mathrm{rms}}^{(k_c)}\propto k_c^{-1/3}$ and $v_{L,\mathrm{rms}}^{(k_c)}\propto k_c^{-0.65}$, respectively.
      In (b), the dashed and dotted lines indicate $T_L^{(k_c)}\propto k_c^{-2/3}$ and $T_L^{(k_c)}\propto k_c^{-0.35}$, respectively.
      The dash-dotted lines represent the Lagrangian integral timescale $T_L$ normalized by $\tau_{\eta,0}$ with colors corresponding to the values of $\mathrm{Wi}_\eta$.}
      \label{fig:lag_bp_vel_statistics}
\end{figure*}

Using $\bm{v}_L^{(k_c)}(t|\bm{x}_0,t_0)$, we quantify the contributions from flows at different scales to the Lagrangian properties.
Figure~\ref{fig:lag_bp_vel_statistics} shows the root-mean-square amplitude $v_{L,\mathrm{rms}}^{(k_c)}=\langle |\bm{v}_L^{(k_c)}|^2\rangle^{1/2}$ and the Lagrangian integral timescale $T_L^{(k_c)}$ of $\bm{v}_L^{(k_c)}(t|\bm{x}_0,t_0)$ as functions of $k_c\eta_0$ for $\mathrm{Re}_{\lambda,0}=218$ with the forcing $\bm{f}^{(\mathrm{D})}$ (Run512D).
Here, $T_L^{(k_c)}$ is defined as 
\begin{equation}
  T_{L}^{(k_c)}=\int_0^\infty C_{L}^{(k_c)}(t)\dd t,
\end{equation}
where the autocorrelation function $C_{L}^{(k_c)}(t)$ is given by
\begin{equation}
  C_{L}^{(k_c)}(t) = \frac{\langle \bm{v}_{L}^{(k_c)}(s+t|\bm{x}_0,t_0)\cdot \bm{v}_{L}^{(k_c)}(s|\bm{x}_0,t_0)\rangle}{\langle |\bm{v}_{L}^{(k_c)}|^2\rangle }.
\end{equation}
We normalize $v_{L,\mathrm{rms}}^{(k_c)}$ and $T_L^{(k_c)}$ by the Kolmogorov velocity $v_{\eta,0}=(\nu_0\overline{\epsilon_0})^{1/4}$ and the Kolmogorov time $\tau_{\eta,0}$ of the corresponding Newtonian case, respectively.
Note that $v_{L,\mathrm{rms}}^{(k_c)}$ yields almost identical information to that of $E(k)$.
For $\mathrm{Wi}_\eta=0$~(i.e., Newtonian turbulence), $v_{L,\mathrm{rms}}^{(k_c)}\propto k_c^{-1/3}$ in the inertial range, which is consistent with the Kolmogorov similarity hypothesis.
For polymer solutions, at fixed $k_c$, $v_{L,\mathrm{rms}}^{(k_c)}$ decreases with increasing $\mathrm{Wi}_\eta$, with the decrease first appearing at large $k_c$ and then extending to smaller $k_c$.
This indicates a reduced contribution of small-scale~(large-$k_c$) flows to the Lagrangian velocity.
For $\mathrm{Wi}_\eta=15$, $v_{L,\mathrm{rms}}^{(k_c)}\propto k_c^{-0.65}$ holds in a narrow range, which is consistent with $E(k)\propto k^{-2.3}$ in Fig.~\ref{fig:ene_spe}.

Next, we examine the characteristic timescales associated with vortices at different scales from the Lagrangian perspective using $T_L^{(k_c)}$.
Figure~\ref{fig:lag_bp_vel_statistics}(b) shows that $T_L^{(k_c)}$ decreases with increasing $k_c$, indicating that smaller-scale vortices contribute to faster temporal variations in the Lagrangian velocity.
For the Newtonian case, the observed scaling $T_L^{(k_c)}\propto k_c^{-2/3}$ is consistent with the prediction of the Kolmogorov similarity hypothesis and with previous observations~\cite{Gotoh1993-qm,Kaneda1999-rl,Matsumoto2021-dg}.
As $\mathrm{Wi}_\eta$ increases, $T_L^{(k_c)}$ increases at large $k_c$ compared with Newtonian turbulence, whereas $T_L^{(k_c)}$ at low $k_c$ remains nearly unchanged.
For $\mathrm{Wi}_\eta=15$, $T_L^{(k_c)}$ at $0.02\lesssim k_c\eta_0\lesssim 0.2$ decreases approximately as $T_L^{(k_c)}\propto k_c^{-0.35}$, consistent with the phenomenological scaling argument $T_L^{(k_c)}\sim (k_c v_{L,\mathrm{rms}}^{(k_c)})^{-1}$ with $v_{L,\mathrm{rms}}^{(k_c)}\propto k_c^{-0.65}$~\cite{Singh2025-ur}.
Thus, polymers not only suppress the amplitude of $\bm{v}_L^{(k_c)}(t|\bm{x}_0,t_0)$ at large $k_c$ but also slow down its fluctuating timescale, while having little effect on $\bm{v}_L^{(k_c)}(t|\bm{x}_0,t_0)$ at low $k_c$.
To understand the $\mathrm{Wi}_\eta$ dependence of $T_L$ shown in Fig.~\ref{fig:lag_acf}(b), we also show $T_L/\tau_{\eta,0}$ for each $\mathrm{Wi}_\eta$ in Fig.~\ref{fig:lag_bp_vel_statistics}(b).
From the viewpoint of our scale-decomposition analysis, the increase in $T_L$ arises from two factors: the increase in $T_L^{(k_c)}$ and the decrease in the amplitude $v_{L,\mathrm{rms}}^{(k_c)}$ of $\bm{v}_L^{(k_c)}(t|\bm{x}_0,t_0)$, both of which occur successively from large to small $k_c$.
As $v_{L,\mathrm{rms}}^{(k_c)}$ at large $k_c$ decreases, the contribution of the rapidly-varying components $\bm{v}_L^{(k_c)}(t|\bm{x}_0,t_0)$ to $\bm{v}_{L}(t|\bm{x}_0,t_0)$ diminishes.
Consequently, for $\mathrm{Wi}_\eta\lesssim 6$, $T_L$ gradually approaches $T_L^{(k_c)}$ of the largest-scale flows~(i.e., at $k_c=\sqrt{2}$).
For $\mathrm{Wi}_\eta=15$, $T_L^{(k_c)}$ slightly increases even at $k_c=\sqrt{2}$ because $\mathrm{De}$ becomes sufficiently close to unity for the largest-scale flows to be modulated, leading to a significant increase in $T_L$~(see Fig.~\ref{fig:lag_acf}).
We have confirmed that the same scenario holds for cases with different Reynolds numbers and different forcing methods, namely Run384D and Run512S~(Appendix~D).
% %   --------------------
%   --------------------

\begin{figure*}
  \centering
      \begin{tabular}{c}
      \begin{minipage}{0.5\hsize}
          \begin{overpic}[width=1\linewidth]{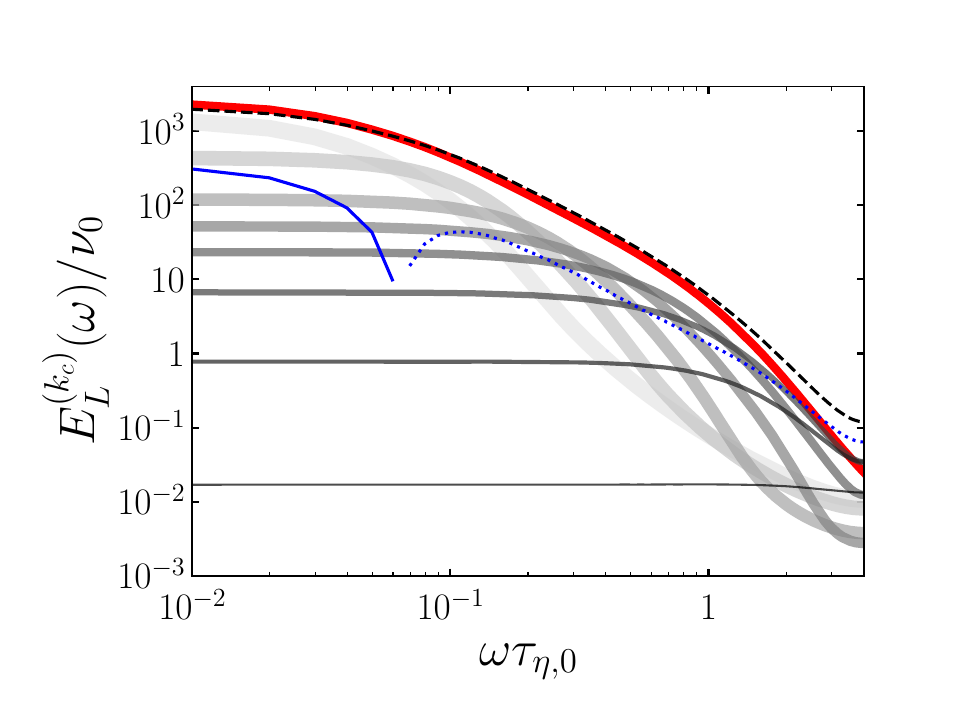}
              \linethickness{3pt}
        \put(5,63){(a)}

          \end{overpic}
      \end{minipage}
      \begin{minipage}{0.5\hsize}
          \begin{overpic}[width=1\linewidth]{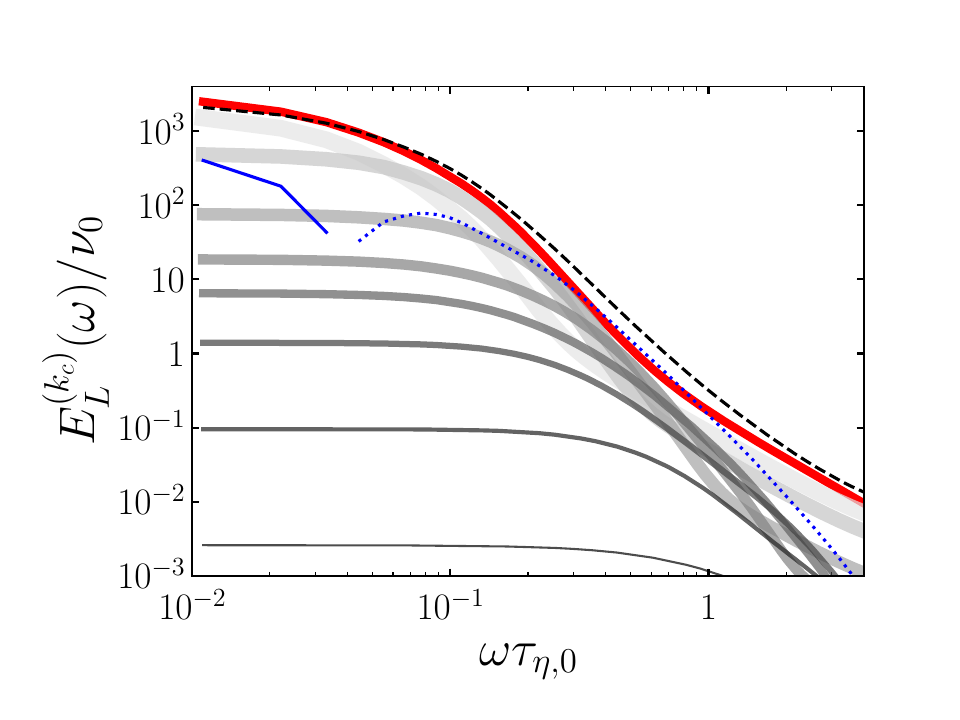}
              \linethickness{3pt}
        \put(5,63){(b)}

          \end{overpic}
      \end{minipage}
      \end{tabular}
  \caption{Power spectral density $E_{L}^{(k_c)}(\omega)$ of the scale-decomposed Lagrangian velocity $\bm{v}_{L}^{(k_c)}(t|\bm{x}_0,t_0)$ for Run512D at (a) $\mathrm{Wi}_\eta=0$~(Newtonian turbulence) and (b) $15$. From thicker~(lighter) to thinner~(darker) lines, $k_c=\sqrt{2},~2\sqrt{2},~4\sqrt{2},~8\sqrt{2},~16\sqrt{2},~32\sqrt{2},~64\sqrt{2}$, and $128\sqrt{2}$. The red line shows $E_{L}(\omega)$. The black dashed line shows the first term on the right-hand side of Eq.~\eqref{eq:Lagrangian_power_spectra_decomposition}. The blue solid and dotted lines indicate the positive and negative values of the second term on the right-hand side of Eq.~\eqref{eq:Lagrangian_power_spectra_decomposition}, respectively.}

      \label{fig:lag_bp_psd}
\end{figure*}

We use the scale-decomposed Lagrangian velocity to clarify which vortex scales are responsible for the modulation behavior of $E_L(\omega)$, including the suppression that successively extends from high to low frequencies and the forcing-dependent behavior at large $\mathrm{Wi}_\eta$~(Fig.~\ref{fig:lag_psd}).
With Eqs.~\eqref{eq:Lagrangian_power_spectra} and \eqref{eq:Lagrangian_velocity_decomposition}, $E_{L}(\omega)$ is expressed as 
\begin{equation}
    E_{L}(\omega) = \sum_{k_c}\frac{2\langle|\hat{\bm{v}}_{L}^{(k_c)}(\omega)|^2\rangle}{T} + \sum_{k_c,k_c^\prime(\neq k_c)}\frac{4\langle\mathop{\mathrm{Re}}\{\hat{\bm{v}}_{L}^{(k_c)}(\omega)\cdot{\hat{\bm{v}}_{L}^{(k_c^\prime)}} (\omega)\}\rangle}{T}.\label{eq:Lagrangian_power_spectra_decomposition}
\end{equation}
Here, $\mathop{\mathrm{Re}}\{\cdot\}$ denotes the real part of a complex number.
Figure~\ref{fig:lag_bp_psd} shows $E_{L}(\omega)$ and each term on the right-hand side of Eq.~\eqref{eq:Lagrangian_power_spectra_decomposition} for $\mathrm{Re}_{\lambda,0}=218$ with the forcing $\bm{f}^{(\mathrm{D})}$ (Run512D) at $\mathrm{Wi}_\eta=0$ and $\mathrm{Wi}_\eta=15$.
Note that we show the absolute value of the second term because its sign can be negative.
Positive and negative values are represented by solid and dotted lines, respectively.
We find that the first term on the right-hand side of Eq.~\eqref{eq:Lagrangian_power_spectra_decomposition} qualitatively reproduces the behavior of $E_{L}(\omega)$, although we cannot completely ignore the second term due to temporal cross-correlations of $\bm{v}_L^{(k_c)}(t|\bm{x}_0,t_0)$ at different wavenumbers. 
While the second term is almost negligible in the inertial range for Newtonian turbulence~[Fig.~\ref{fig:lag_bp_psd}(a)], its magnitude becomes comparable to that of $E_{L}(\omega)$ within a high-frequency range for $\mathrm{Wi}_\eta=15$~~[Fig.~\ref{fig:lag_bp_psd}(b)], resulting in a quantitative discrepancy between $E_{L}(\omega)$ and the first term on the right-hand side of Eq.~\eqref{eq:Lagrangian_power_spectra_decomposition}.
Keeping this limitation in mind, we focus on the first term on the right-hand side of Eq.~\eqref{eq:Lagrangian_power_spectra_decomposition} to relate the $\mathrm{Wi}_\eta$ dependence of $E_{L}(\omega)$ to the coherent vortices at different length scales.
We define the contribution $E_{L}^{(k_c)}(\omega)$ of the flow at wavenumber $k_c$ to $E_{L}(\omega)$ as
\begin{equation}
    E_{L}^{(k_c)}(\omega) = \frac{2\langle|\hat{\bm{v}}_{L}^{(k_c)}(\omega)|^2\rangle}{T}.
\end{equation}
We show $E_{L}^{(k_c)}(\omega)$ for different $k_c$ in Fig.~\ref{fig:lag_bp_psd} with gray lines, indicating that $E_{L}(\omega)$ is composed of contributions from vortices of different sizes.
In the following, we demonstrate how the contribution $E_{L}^{(k_c)}(\omega)$ from each scale to $E_L(\omega)$ varies with $\mathrm{Wi}_\eta$, thereby elucidating the modulation mechanism of $E_L(\omega)$.
% %   --------------------
%   --------------------

\begin{figure}
  \centering
  \begin{overpic}[width=1\linewidth]{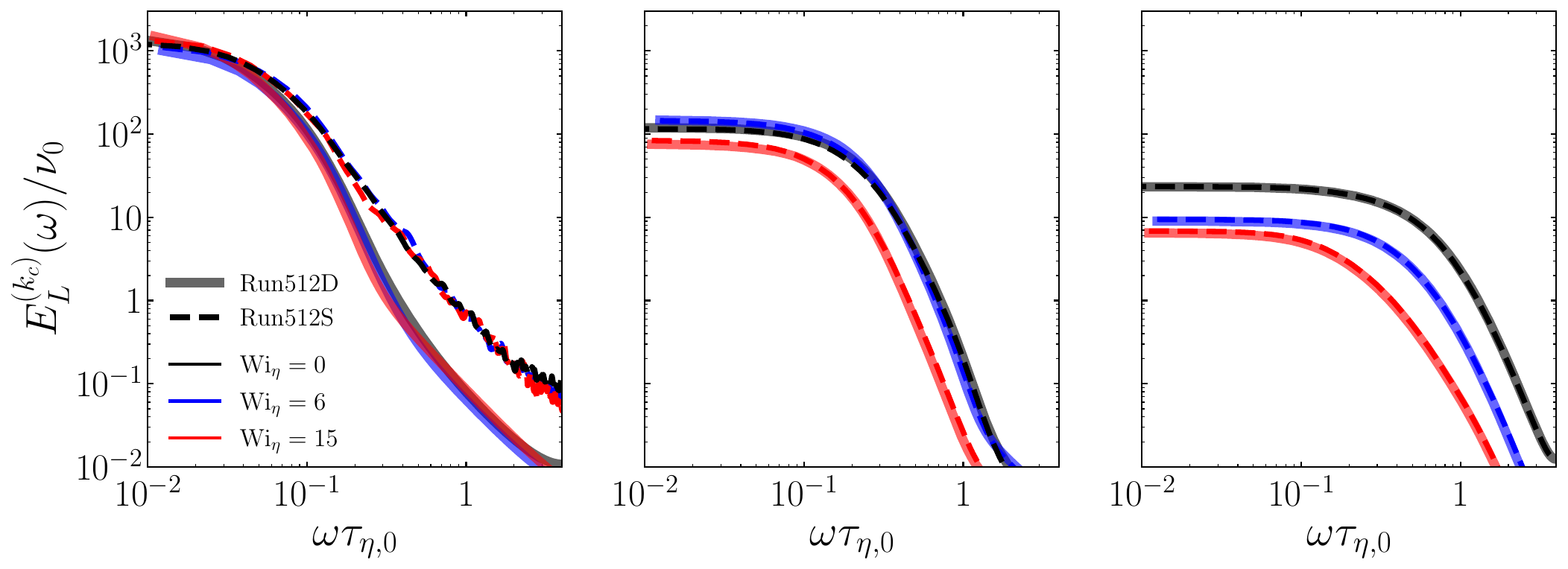} 
    \put(4,35){(a)}
    \put(37.5,35){(b)}
    \put(69,35){(c)}
  \end{overpic}
  \caption{Power spectral density $E_L^{(k_c)}(\omega)$ of the scale-decomposed Lagrangian velocity $\bm{v}_L^{(k_c)}(t|\bm{x}_0,t_0)$ at (a) $k_c\eta_0=0.0080$, (b) $0.032$, and (c) $0.13$ for Run512D~(solid lines) and Run512S~(dashed lines). Different colors correspond to different values of $\mathrm{Wi}_\eta$: black, $\mathrm{Wi}_\eta=0$; blue, $6$; red, $15$.}
  \label{fig:lag_bp_psd_kc}
\end{figure}

We explain the $\mathrm{Wi}_\eta$ dependence of $E_L(\omega)$ using the contributions $E_{L}^{(k_c)}(\omega)$ from each scale to $E_L(\omega)$.
Here, we focus on three wavenumbers $k_c=\sqrt{2}$, $4\sqrt{2}$, and $16\sqrt{2}$, which, for Run512D and Run512S, correspond to $k_c\eta_0=0.0080$, $0.032$, and $0.13$, respectively.
As already demonstrated in Fig.~\ref{fig:snapshot}, vortices extracted by band-pass filtering with different $k_c$ exist in a hierarchical manner in Newtonian turbulence, whereas small-scale vortices at $k_c=16\sqrt{2}$ disappear at $\mathrm{Wi}_\eta=6$, and those at $k_c=4\sqrt{2}$ are also attenuated at $\mathrm{Wi}_\eta=15$.
Figure~\ref{fig:lag_bp_psd_kc} shows $E_{L}^{(k_c)}(\omega)$ for the three wavenumbers at $\mathrm{Wi}_\eta=0$~(black), $6$~(blue), and $15$~(red).
Each panel shows the results for fixed $k_c\eta_0$.
We first focus on the results for $\mathrm{Re}_{\lambda,0}=218$ with the forcing $\bm{f}^{(\mathrm{D})}$ (Run512D), shown by the thick solid lines.
For $\mathrm{Wi}_\eta=6$~(blue solid lines), Fig.~\ref{fig:lag_bp_psd_kc}(c) shows that $E_{L}^{(k_c)}(\omega)$ at $k_c\eta_0=0.13$ declines compared with the Newtonian case, as expected from the visualization of vortices~[Fig.~\ref{fig:snapshot}(b2)] and the energy spectrum~(Fig.~\ref{fig:ene_spe}).
In contrast, Figs.~\ref{fig:lag_bp_psd_kc}(a) and (b) demonstrate that $E_{L}^{(k_c)}(\omega)$ at $k_c\eta_0=0.0080$ and $0.032$ is almost identical to its Newtonian counterpart in both magnitude and spectral shape over the entire frequency range.
Our observations indicate that, at length scales where the Eulerian statistics are insensitive to polymer effects, the corresponding contribution to the Lagrangian velocity dynamics is also insensitive to polymer effects.
For $\mathrm{Wi}_\eta=15$~(red solid lines), $E_{L}^{(k_c)}(\omega)$ at $k_c\eta_0=0.032$ also declines relative to the Newtonian case, while that at $k_c\eta_0=0.0080$ shows only a slight change in spectral shape.
Thus, Fig.~\ref{fig:lag_bp_psd_kc} reveals that even from a Lagrangian perspective, polymer effects become evident successively from smaller-scale vortices as $\mathrm{Wi}_\eta$ increases.

We also examine whether the modulation mechanism of $E_L(\omega)$ depends on the forcing method by comparing $E_{L}^{(k_c)}(\omega)$ obtained with the forcings $\bm{f}^{(\mathrm{S})}$ and $\bm{f}^{(\mathrm{D})}$, corresponding to Run512S and Run512D, respectively.
Figure~\ref{fig:lag_bp_psd_kc} demonstrates that, for a given $\mathrm{Wi}_\eta$, $E_{L}^{(k_c)}(\omega)$ is essentially identical between Run512S (dashed lines) and Run512D (solid lines), except at the lowest wavenumber, $k_c\eta_0=0.0080$.
Because this wavenumber corresponds to the forcing scale, the spectral shape there differs due to the forcing method.

Therefore, the modulation mechanism of $E_L(\omega)$ by polymers is summarized as follows.
As the polymer relaxation time increases, turbulent vortices are attenuated successively from smaller scales because smaller-scale vortices have shorter characteristic timescales, or equivalently higher strain rates, and therefore interact more strongly with polymers~\cite{lumley1969drag,Tabor1986-yi}.
This attenuation leads to a decrease in $E_L(\omega)$ in the high-frequency range~(Fig.~\ref{fig:lag_psd}) because smaller-scale vortices induce higher-frequency fluctuations of the Lagrangian velocity~(Fig.~\ref{fig:lag_bp_psd}).
In contrast, large-scale vortices are almost insensitive to polymer effects, not only from an Eulerian perspective but also from a Lagrangian perspective~(Fig.~\ref{fig:lag_bp_psd_kc}).
Thus, $E_L(\omega)$ at low $\omega$ remains close to its Newtonian counterpart because the low-frequency fluctuations are primarily generated by these large-scale vortices.
%   --------------------
%   --------------------

\begin{figure}
  \centering
  \begin{overpic}[width=0.5\linewidth]{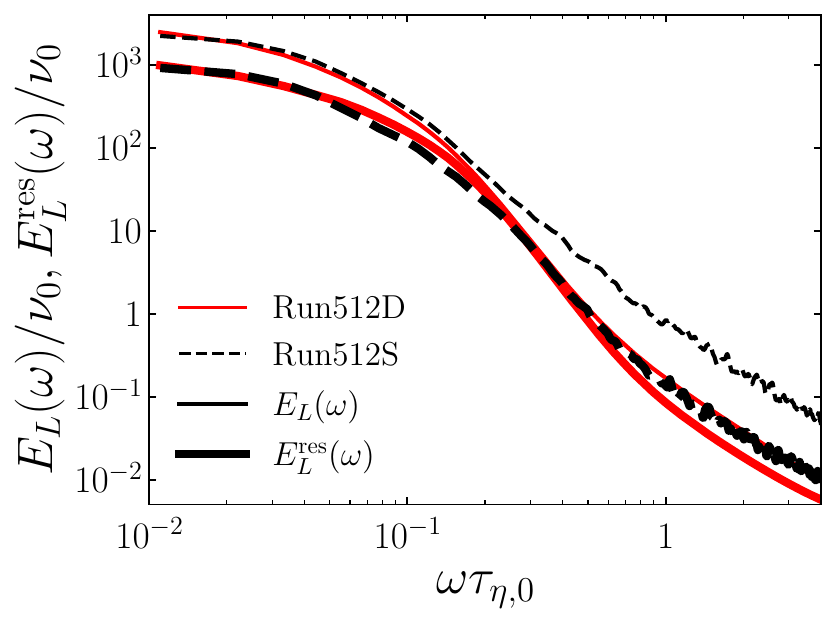} 
  \end{overpic}
  \caption{Power spectral density $E_L(\omega)$~(thin lines) of the Lagrangian velocity $\bm{v}_{L}(t|\bm{x}_0,t_0)$ and the residual power spectral density $E_L^\mathrm{res}(\omega)$~(thick lines) obtained by subtracting the contribution $E_L^{(\sqrt{2})}(\omega)$ at the forcing scale. Results are shown for Run512D~($\mathrm{Re}_{\lambda,0}=218$, $\bm{f}^{(\mathrm{D})}$; red solid lines) and Run512S~($\mathrm{Re}_{\lambda,0}=225$, $\bm{f}^{(\mathrm{S})}$; black dashed lines) at $\mathrm{Wi}_\eta=15$.}

  \label{fig:lag_psd_wo_largest}
\end{figure}

It is worth emphasizing that our scale-decomposition analysis also explains the $\bm{f}$ dependence of $E_L(\omega)$ at large $\mathrm{Wi}_\eta$~(Fig.~\ref{fig:lag_psd}).
Here, we assume a scenario in which the largest-scale flows driven by the external force mask the underlying universality of $E_L(\omega)$.
Figure~\ref{fig:lag_psd_wo_largest} shows $E_L(\omega)$~(thin lines) and the residual power spectral density $E_L^\mathrm{res}(\omega)=E_L(\omega)-E_{L}^{(\sqrt{2})}(\omega)$~(thick lines) obtained by subtracting the contribution $E_{L}^{(\sqrt{2})}(\omega)$ at the forcing scale.
Here, we consider $\mathrm{Wi}_\eta=15$.
As shown in Fig.~\ref{fig:lag_psd}, $E_L(\omega)$ at large $\omega$ depends on $\bm{f}$ even at nearly the same $\mathrm{Re}_{\lambda,0}$ and $\mathrm{Wi}_\eta$.
However, we observe little difference in $E_L^\mathrm{res}(\omega)$ after removing the contribution from the largest-scale flows.
This collapse indicates that the apparent non-universality of $E_L(\omega)$ in the high-frequency regime does not originate from the small-scale flows, but rather from the largest-scale flows driven by the external force.
Consequently, once the forcing-dependent largest-scale contribution is removed, $E_L^\mathrm{res}(\omega)$ exhibits common behavior even at large $\mathrm{Wi}_\eta$, at least for the forcing methods considered here.

%   --------------------
%   --------------------
\section{Conclusions}
\label{sec:conclusion}

Using the FENE-P model, we have investigated the Lagrangian velocity statistics of turbulence in dilute polymer solutions by varying the parameters $\mathrm{Re}_{\lambda,0}$, $\bm{f}$, and $\mathrm{Wi}_\eta$.
We have demonstrated that as $\mathrm{Wi}_\eta$ increases, the suppression of the power spectral density $E_L(\omega)$ of the Lagrangian velocity extends from large to small $\omega$~(Fig.~\ref{fig:lag_psd}).
Additionally, the Lagrangian integral timescale $T_L$ increases with $\mathrm{Wi}_\eta$~(Fig.~\ref{fig:lag_acf}).

To reveal the physical mechanism underlying the polymer-induced modulation of the Lagrangian velocity statistics, we have conducted the scale-decomposition analysis of the Lagrangian velocity.
We have introduced the contribution $\bm{v}_L^{(k_c)}(t|\bm{x}_0,t_0)$ to the Lagrangian velocity from the flow at wavenumber $k_c$ by applying a bandpass filter to the velocity field $\bm{u}(\bm{x},t)$~(Fig.~\ref{fig:lag_bp_vel}).
Our scale-decomposition analysis has demonstrated that the observed increase in $T_L$ with $\mathrm{Wi}_\eta$ arises from a decrease in the amplitude $v_{L,\mathrm{rms}}^{(k_c)}$ of $\bm{v}_L^{(k_c)}(t|\bm{x}_0,t_0)$ at large $k_c$ and an increase in the corresponding integral timescale $T_L^{(k_c)}$~(Fig.~\ref{fig:lag_bp_vel_statistics}).

The scale-decomposed Lagrangian velocity enables us to decompose $E_L(\omega)$ into the contributions $E_L^{(k_c)}(\omega)$ from different scales~(Fig.~\ref{fig:lag_bp_psd}).
Comparing $E_L^{(k_c)}(\omega)$ at different $\mathrm{Wi}_\eta$ has provided insight into the modulation mechanism of $E_L(\omega)$ in terms of the hierarchy of coherent vortices~(Fig.~\ref{fig:lag_bp_psd_kc}).
As $\mathrm{Wi}_\eta$ increases, the contribution $E_L^{(k_c)}(\omega)$ from small-scale vortices decreases due to their attenuation, leading to a decrease in $E_L(\omega)$ in the high-frequency range.
In contrast, $E_L^{(k_c)}(\omega)$ associated with large-scale vortices remains unchanged, indicating that they are insensitive to polymer effects from both the Eulerian and Lagrangian perspectives.
Consequently, $E_L(\omega)$ in the low-frequency range is nearly identical to that of Newtonian turbulence.
Therefore, we have concluded that the suppression of the Lagrangian velocity spectra extends from high to low frequencies, in the same way that the suppression of the Eulerian energy spectra extends from high to low wavenumbers; the essential underlying mechanism is the successive polymer-induced suppression of coherent vortices from smaller to larger scales.

At first glance, the apparent similarity in the suppression behavior of $E_L(\omega)$ and $E(k)$ may seem trivial without using scale-decomposition analysis.
However, as reported for Newtonian turbulence, Lagrangian statistics are inherently more sensitive than Eulerian statistics to the forcing-dependent nature of the largest-scale flows~\cite{Koide2025-lz}.
Indeed, the present study has demonstrated that, at large $\mathrm{Wi}_\eta$, the largest-scale flows are responsible for the forcing-dependent differences in $E_L(\omega)$ even at high frequencies~(Fig.~\ref{fig:lag_psd_wo_largest}).
Thus, scale-decomposition analysis is essential for identifying the physical origin of the suppression behavior in the Lagrangian velocity spectra.

The present study has paved the way for understanding turbulence modulation by polymer additives from a Lagrangian perspective by systematically demonstrating how polymers modulate the Lagrangian velocity statistics and elucidating the role of different-scale vortices in this modulation through a scale-decomposition approach.
Although the Lagrangian velocity is directly related to the translational motion of the polymer center of mass, velocity gradients are more directly relevant to the stretching and orientation of polymer chains.
Thus, an important direction for future work is to apply our approach to the velocity-gradient tensor along Lagrangian trajectories to reveal the dynamical interactions between turbulent vortices and polymers.

\ack
The present study was supported by JSPS Grants-in-Aid for Scientific Research (24KJ0109, 25K01158, and 26K17304) and Grants-in-Aid for Transformative Research Areas~(26H00389). 
The simulations were mainly conducted on the supercomputer ``Flow'' at the Information Technology Center, Nagoya University, under the HPCI Research Project (hp250196).
Some of the simulations were conducted under the auspices of ``Collaborative Research Project on Computer Science with High-Performance Computing in Nagoya University".
% \end{acknowledgments}
\appendix
\section*{\label{sec:Appendix_grid}Appendix A: Validation of grid resolution}
\renewcommand{\theequation}{A\arabic{equation} }
\renewcommand{\thefigure}{A\arabic{figure}}
\setcounter{equation}{0}
In this appendix, we demonstrate that the spatial resolution is sufficient for evaluating the statistics presented in the main text.
Figure~\ref{fig:grid} shows the energy spectrum $E(k)$ and the power spectral density $E_L(\omega)$ of the Lagrangian velocity for $N^3=384^3$ and $\mathrm{Re}_{\lambda,0}=135$ with the forcing $\bm{f}^{(\mathrm{D})}$~(Run384D) at $\mathrm{Wi}_\eta=6$~(blue solid lines).
We also show the results obtained with a higher resolution~($N^3=512^3$) for the same parameters~(yellow dashed lines).
The results for $N^3=384^3$ and $512^3$ are in good agreement, although the power-law decay of $E(k)$ at $k\eta_0\gtrsim 0.5$, reported by a previous extensive DNS study~\cite{Singh2025-ur}, extends to larger $k$ with increasing $N^3$.
Therefore, the spatial resolution used in this study is sufficient to justify our conclusions.
% %   --------------------
\begin{figure*}
  \centering
      \begin{tabular}{c}
      \begin{minipage}{0.5\hsize}
          \begin{overpic}[width=1\linewidth]{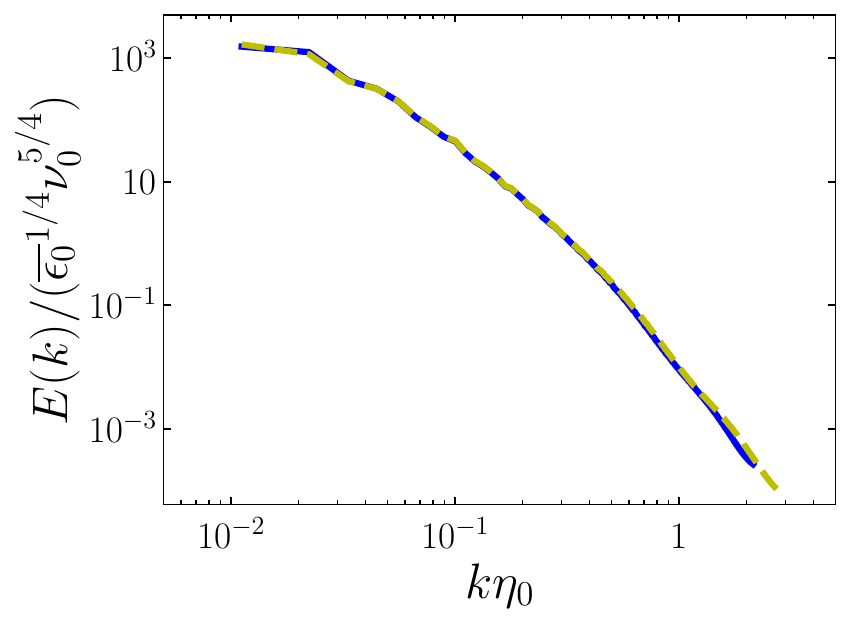}
              \linethickness{3pt}
        \put(1,70){(a)}

          \end{overpic}
      \end{minipage}
      \begin{minipage}{0.5\hsize}
          \begin{overpic}[width=1\linewidth]{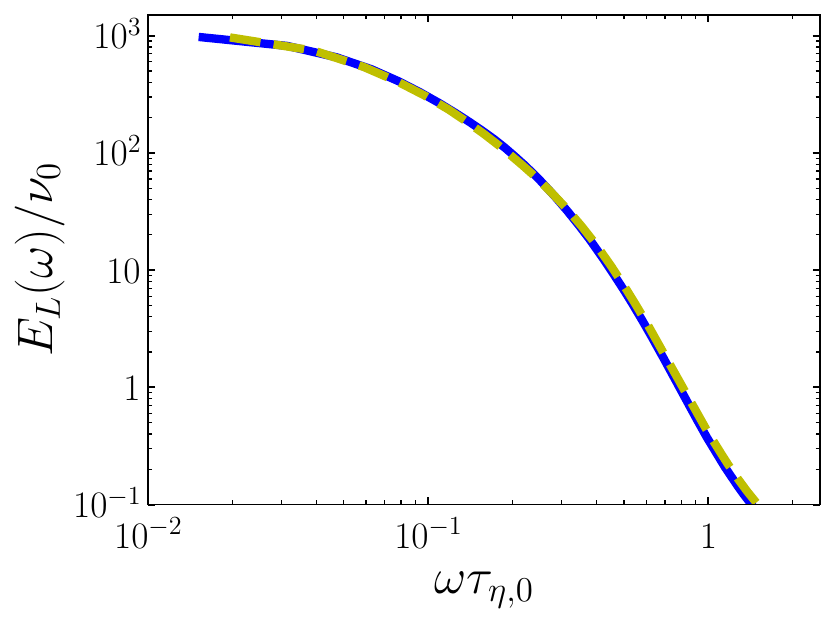}
              \linethickness{3pt}
        \put(1,70){(b)}

          \end{overpic}
      \end{minipage}
      \end{tabular}
  \caption{(a) Energy spectrum $E(k)$ and (b) power spectral density $E_{L}(\omega)$ of the Lagrangian velocity for Run384D at $\mathrm{Wi}_\eta=6$~(blue solid lines). The yellow dashed lines show the results obtained with a higher-resolution $N^3=512^3$.  Both $E(k)$ and $E_{L}(\omega)$ are normalized using $\overline{\epsilon_0}$ and $\nu_0$.}

      \label{fig:grid}
\end{figure*}
%   --------------------
\section*{\label{sec:Appendix_euler}Appendix B: Power spectral density of the Eulerian velocity}
\renewcommand{\theequation}{B\arabic{equation} }
\renewcommand{\thefigure}{B\arabic{figure}}
\setcounter{figure}{0}

\setcounter{equation}{0}

\begin{figure}
  \centering
  \begin{overpic}[width=0.5\linewidth]{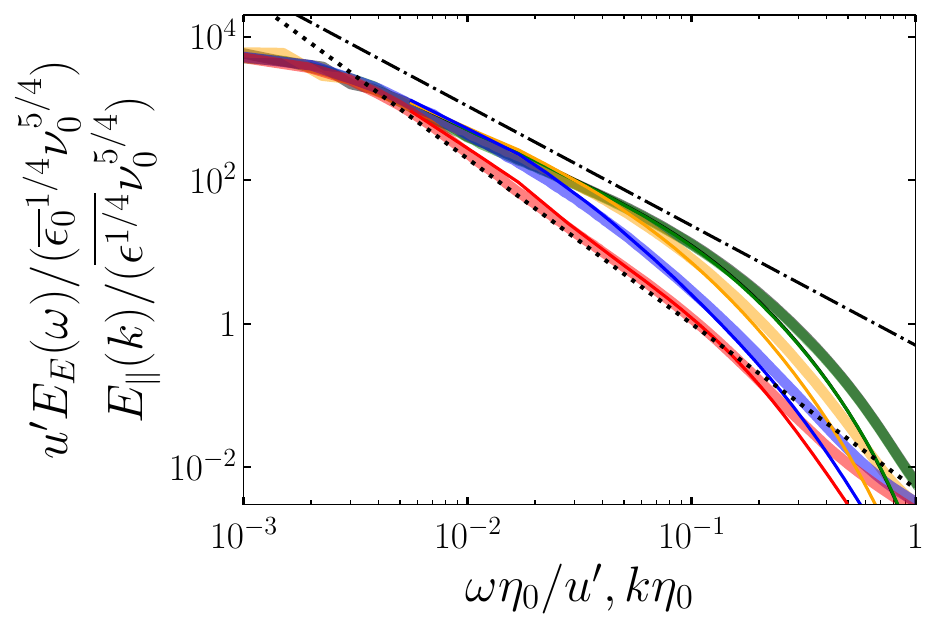} 
  \end{overpic}
  \caption{Power spectral density $E_E(\omega)$~(thick lines) of the Eulerian velocity $\bm{u}_{E}(t|\bm{x})$ and one-dimensional longitudinal energy spectrum $E_\parallel(k)$~(thin lines) for $\mathrm{Re}_{\lambda,0}=218$ with the forcing $\bm{f}^{(\mathrm{D})}$ (Run512D) at $\mathrm{Wi}_\eta=0$~(black), $1$~(green), $3$~(orange), $6$~(blue), and $15$~(red). The black dash-dotted and dotted lines indicate $E_E(\omega)\propto \omega^{-5/3}$ and $E_E(\omega)\propto \omega^{-2.3}$, respectively.}

  \label{fig:euler_psd}
\end{figure}

The present study has focused on the time series of the Lagrangian velocity.
It is worth examining the Eulerian velocity measured at a fixed position, which is likewise a time series, to compare the Lagrangian and Eulerian viewpoints.
For a given time series of the Eulerian velocity $\bm{u}_{E}(t|\bm{x})$ at a fixed position $\bm{x}$, $E_{E}(\omega)$ is defined as
\begin{equation}
    E_{E}(\omega) = \frac{1}{3}\frac{\overline{|\hat{\bm{u}}_{E}(\omega)|^2}}{T}, \label{eq:Euler_psd}
\end{equation}
where $\hat{\bm{u}}_{E}(\omega)$ is the Fourier coefficient of $\bm{u}_{E}(t|\bm{x})$, $T$ is the length of the time series of $\bm{u}_{E}(t|\bm{x})$, and $\overline{(\cdot)}$ denotes spatial averaging.
Note that $E_{E}(\omega)$ defined by Eq.~\eqref{eq:Euler_psd} satisfies 
\begin{equation}
    \int_0^\infty E_{E}(\omega) \dd\omega = \frac{1}{6}\langle |\bm{u}_{E}|^2\rangle,%\lim_{T\to \infty}\frac{1}{T}\int_0^T v_{L,\alpha}^2dt.
\end{equation}
which is suitable for comparison with the one-dimensional energy spectrum below.
Figure~\ref{fig:euler_psd} shows $E_{E}(\omega)$ for $\mathrm{Re}_{\lambda,0}=218$ with the forcing $\bm{f}^{(\mathrm{D})}$ (Run512D) at different values of $\mathrm{Wi}_\eta$.
For comparison, we also show the one-dimensional longitudinal energy spectrum $E_\parallel(k)$.
Here, using $E(k)$, we compute $E_\parallel(k)$ as
\begin{equation}
  E_\parallel(k) = \frac{1}{2}\int_{k}^\infty \frac{E(k^\prime)}{k^\prime}\left(1-\frac{k^2}{k^{\prime 2}}\right)\dd k.
\end{equation}
To compare $E_{E}(\omega)$ with $E_\parallel(k)$, we show $u^\prime E_E(\omega)/(\overline{\epsilon_0}^{1/4}\nu_0^{5/4})$ as a function of $\omega\eta_0/u^\prime$, where $u^\prime$ is evaluated separately for each case as the root-mean-square of $\bm{u}_{E}(t|\bm{x})$.
Here, we assume that the sweeping of small-scale vortices by large-scale flows has a dominant influence on $\bm{u}_{E}(t|\bm{x})$, which leads to the relations $k\sim \omega/u^\prime$ and $E_\parallel(k)\sim u^\prime E_{E}(\omega)$.
We find that for each $\mathrm{Wi}_\eta$, $u^\prime E_E(\omega)/(\overline{\epsilon_0}^{1/4}\nu_0^{5/4})$ and $E_\parallel(k)/(\overline{\epsilon_0}^{1/4}\nu_0^{5/4})$ nearly collapse for $\omega\eta_0/u^{\prime}$~(or $k\eta_0$)$\lesssim 0.2$.
This collapse suggests that the Taylor hypothesis~\cite{Taylor1938-bu,Tennekes1975-gz} remains approximately valid in this range, even for turbulent flows with polymer additives without a mean flow.
Thus, $E_E(\omega)$ mainly reflects spatial information about turbulence due to the sweeping of small-scale vortices by large-scale flows, as reported in our previous work for Newtonian turbulence~\cite{Koide2025-lz}.
However, whether this correspondence persists at larger $\omega\eta_0/u^{\prime}$~(or $k\eta_0$), where elastic effects become significant, requires careful examination, as has been done for elastic turbulence at low Reynolds numbers~\cite{Foggi-Rota2026-tw}.
%   --------------------
%   --------------------

\section*{\label{sec:Appendix_beta_Lp}Appendix C: Effect of the concentration and maximum extensibility of polymers}
\renewcommand{\theequation}{C\arabic{equation} }
\renewcommand{\thefigure}{C\arabic{figure}}
\setcounter{figure}{0}
\setcounter{equation}{0}

\begin{figure*}
  \centering
      \begin{tabular}{c}
      \begin{minipage}{0.5\hsize}
          \begin{overpic}[width=1\linewidth]{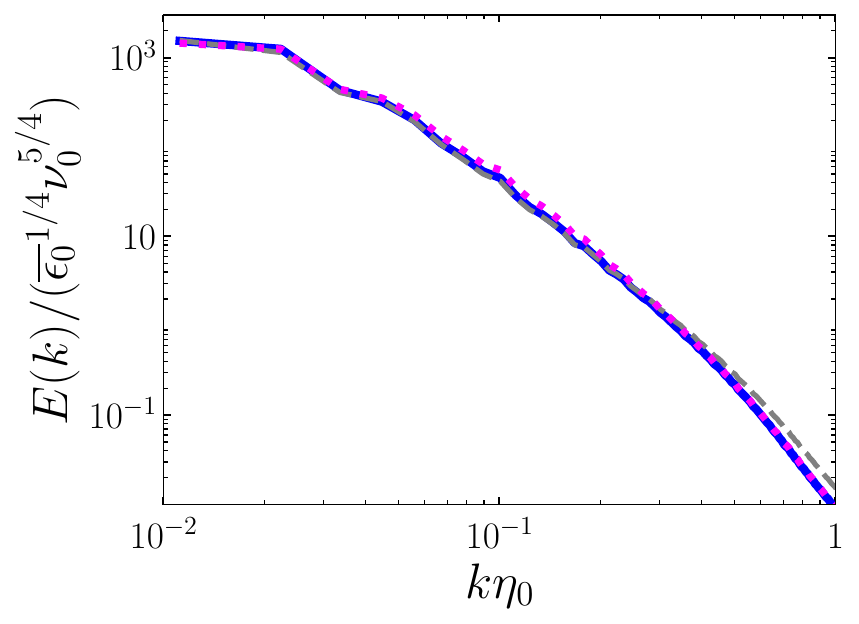}
              \linethickness{3pt}
        \put(1,70){(a)}

          \end{overpic}
      \end{minipage}
      \begin{minipage}{0.5\hsize}
          \begin{overpic}[width=1\linewidth]{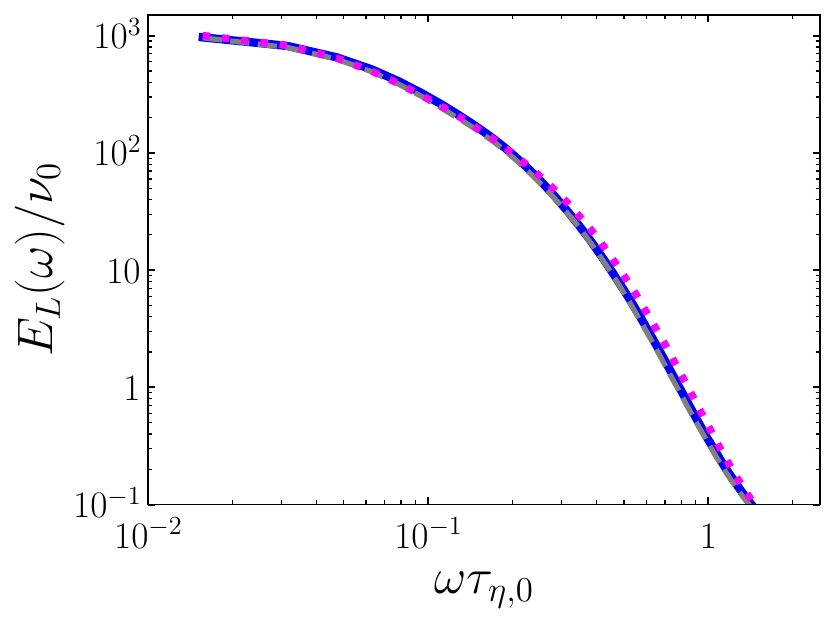}
              \linethickness{3pt}
        \put(1,70){(b)}

          \end{overpic}
      \end{minipage}
      \end{tabular}
  \caption{(a) Energy spectrum $E(k)$ and (b) power spectral density $E_{L}(\omega)$ of the Lagrangian velocity for $\mathrm{Re}_{\lambda,0}=135$ with the forcing $\bm{f}^{(\mathrm{D})}$~(Run384D) at fixed $\mathrm{Wi}_\eta=6$. Different lines denote different values of $(\beta,L_\mathrm{p})$: blue solid line, $(\beta,L_\mathrm{p})=(0.9,1000)$; gray dashed line,~$(0.6,1000)$; magenta dotted line, $(0.9,100)$. Both $E(k)$ and $E_{L}(\omega)$ are normalized using $\overline{\epsilon_0}$ and $\nu_0$.}

      \label{fig:beta_Lp}
\end{figure*}

In the main text, we have fixed $\beta(=0.9)$ and $L_\mathrm{p}(=1000)$ to focus on the effect of the polymer relaxation time.
This appendix briefly examines the effects of $\beta$ and $L_\mathrm{p}$.
Figure~\ref{fig:beta_Lp} shows $E(k)$ and $E_L(\omega)$ for $\mathrm{Re}_{\lambda,0}=135$ with the forcing $\bm{f}^{(\mathrm{D})}$~(Run384D) at fixed $\mathrm{Wi}_\eta(=6)$.
We conduct additional DNS for $(\beta,L_\mathrm{p})=(0.6,1000)$ and $(0.9,100)$.
Figure~\ref{fig:beta_Lp}(a) demonstrates that within the range considered, $\beta$ and $L_\mathrm{p}$ have little effect on $E(k)$.
Accordingly, $E_L(\omega)$ for different values of $\beta$ and $L_\mathrm{p}$ also collapses onto a single function for fixed $\mathrm{Wi}_\eta$.
As one would expect, $E(k)$ and $E_L(\omega)$ may vary when $\beta$ approaches $0$ or $1$, or when $L_\mathrm{p}$ is small enough.
However, a systematic investigation with respect to $\beta$ and $L_\mathrm{p}$ is beyond the scope of this study. 
In this appendix, we demonstrate that variations in $\beta$ and $L_\mathrm{p}$ that keep $E(k)$~(i.e., the spatial structure of turbulence) almost unchanged have only a negligible effect on $E_L(\omega)$.
% %   --------------------
%   --------------------
\section*{\label{sec:Appendix_lag_bp}Appendix D: Lagrangian integral timescale of the scale-decomposed Lagrangian velocity for different forcing methods and Reynolds numbers}
\renewcommand{\theequation}{D\arabic{equation} }
\renewcommand{\thefigure}{D\arabic{figure}}
\setcounter{figure}{0}
\setcounter{equation}{0}
In the main text, we have explained the physical mechanism underlying the increase in $T_L$ with $\mathrm{Wi}_\eta$ using a scale-decomposition analysis~(Fig.~\ref{fig:lag_bp_vel_statistics}).
We have focused on the modulation of $\bm{v}_L^{(k_c)}$ from the perspectives of both its amplitude ${v}_{L,\mathrm{rms}}^{(k_c)}$ and Lagrangian integral timescale $T_L^{(k_c)}$.
Specifically, as $\mathrm{Wi}_\eta$ increases, ${v}_{L,\mathrm{rms}}^{(k_c)}$ decreases while $T_L^{(k_c)}$ increases, both successively from large to small $k_c$, thus leading to the increase in $T_L$.
This appendix demonstrates that this scenario holds for Run512S and Run384D, which differ in forcing method and Reynolds number, respectively.
Since ${v}_{L,\mathrm{rms}}^{(k_c)}$ provides essentially the same information as $E(k)$, as noted in the main text, the attenuation of ${v}_{L,\mathrm{rms}}^{(k_c)}$ for Run512S and Run384D is evident from $E(k)$~(see Fig.~\ref{fig:ene_spe}). 
Thus, we focus on the $\mathrm{Wi}_\eta$ dependence of $T_L^{(k_c)}$.
Figure~\ref{fig:lag_bp_time} shows $T_L^{(k_c)}$ as a function of $k_c\eta_0$ for Run512S and Run384D at various $\mathrm{Wi}_\eta$.
As in Run512D, an increase in $T_L^{(k_c)}$ extends from large to small $k_c$ as $\mathrm{Wi}_\eta$ increases.
Therefore, we conclude that the observed increase in $T_L$~[Fig.~\ref{fig:lag_acf}(b)] is attributed to both the decrease in ${v}_{L,\mathrm{rms}}^{(k_c)}$ and the increase in $T_L^{(k_c)}$.

% %   --------------------
\begin{figure*}
  \centering
      \begin{tabular}{c}
      \begin{minipage}{0.5\hsize}
          \begin{overpic}[width=1\linewidth]{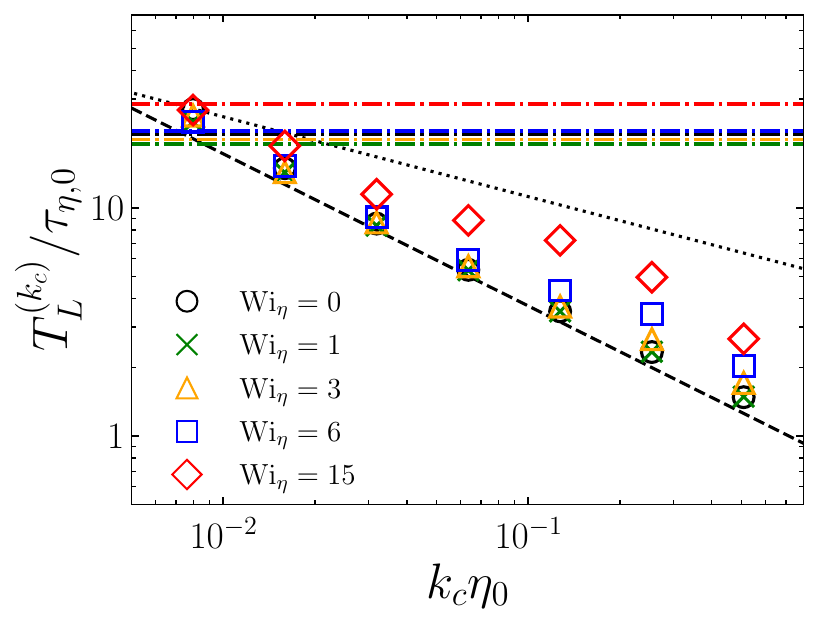}
              \linethickness{3pt}
        \put(1,70){(a)}

          \end{overpic}
      \end{minipage}
      \begin{minipage}{0.5\hsize}
          \begin{overpic}[width=1\linewidth]{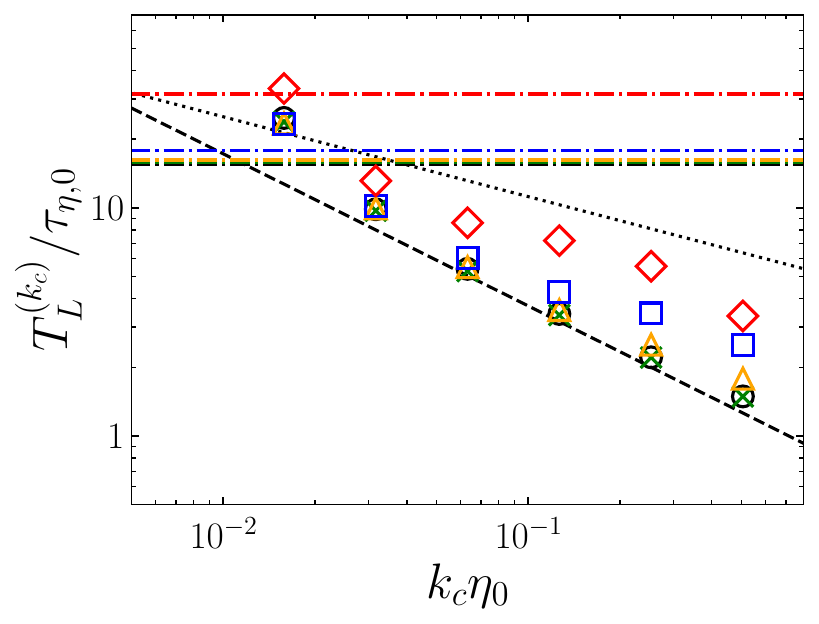}
              \linethickness{3pt}
        \put(1,70){(b)}

          \end{overpic}
      \end{minipage}
      \end{tabular}
  \caption{Lagrangian integral timescale $T_L^{(k_c)}$ of the scale-decomposed Lagrangian velocity $\bm{v}_{L}^{(k_c)}(t|\bm{x}_0,t_0)$ as a function of $k_c\eta_0$ for (a) $\mathrm{Re}_{\lambda,0}=225$ with the forcing $\bm{f}^{(\mathrm{S})}$~(Run512S) and (b) $\mathrm{Re}_{\lambda,0}=135$ with the forcing $\bm{f}^{(\mathrm{D})}$~(Run384D) at various values of $\mathrm{Wi}_\eta$.
  The dashed and dotted lines indicate $T_L^{(k_c)}\propto k_c^{-2/3}$ and $T_L^{(k_c)}\propto k_c^{-0.35}$, respectively.
  The dash-dotted lines represent the Lagrangian integral timescale $T_L$ normalized by $\tau_{\eta,0}$ with colors corresponding to the values of $\mathrm{Wi}_\eta$.
  }

      \label{fig:lag_bp_time}
\end{figure*}
%   --------------------
\providecommand{\newblock}{}

\end{document}